\documentclass[aps,prd,10pt,nofootinbib,twocolumn,eqsecnum,showpacs,showkeys,superscriptaddress,preprintnumbers,altaffilletter]{revtex4-1}
\usepackage{amsmath}
\allowdisplaybreaks

\usepackage{multirow}

\usepackage{soul}
\usepackage{graphicx}
\usepackage{amsmath}
\usepackage{multirow}
\usepackage{dcolumn}
\usepackage{amssymb}
\usepackage{amsfonts}
\usepackage{amsbsy}
\usepackage{xcolor}
\usepackage{rotating}
\usepackage[english]{babel}
\usepackage{float}
\usepackage{orcidlink}
\usepackage{tabularray}

\usepackage{hyperref}
\hypersetup{
    colorlinks=true,
    linkcolor=magenta,
    filecolor=magenta,
    urlcolor=magenta,
    citecolor=magenta
}

\usepackage[normalem]{ulem}
\usepackage{comment}
\usepackage{caption}
\usepackage{booktabs}
\usepackage{threeparttable}
\usepackage{tablefootnote}
\usepackage{enumitem}

\usepackage{booktabs}
\usepackage{threeparttable}

\usepackage{subcaption}

\newcommand{\be}{\begin{equation}}
\newcommand{\ee}{\end{equation}}
\newcommand{\bea}{\begin{eqnarray}}
\newcommand{\eea}{\end{eqnarray}}
\newcommand{\beal}{\begin{aligned}}
\newcommand{\eeal}{\end{aligned}}
\newcommand{\bi}{\begin{itemize}}
\newcommand{\ei}{\end{itemize}}

\newcommand{\tightdoublehline}{%
  \noalign{\vskip 0.3pt}\hline
  \noalign{\vskip 0.3pt}\hline
  \noalign{\vskip 0.6pt}%
}

\usepackage{xurl}
\begin{document}

\title{Exploring Non-minimal coupling using ultra-diffuse galaxies}

\author{Saboura Zamani\,\orcidlink{0009-0004-3201-9483}}
\email{saboura.zamani@phd.usz.edu.pl}
\affiliation{Institute of Physics, University of Szczecin, Wielkopolska 15, 70-451 Szczecin, Poland}
\author{Vincenzo Salzano\, \orcidlink{0000-0002-4905-1541}}
\email{vincenzo.salzano@usz.edu.pl}
\affiliation{Institute of Physics, University of Szczecin, Wielkopolska 15, 70-451 Szczecin, Poland}
\author{Dario Bettoni\, \orcidlink{0000-0002-0176-5537}}
\email{dbet@unileon.es}
\affiliation{Departamento de Matemáticas, Universidad de León,
Escuela de Ingenierías Industrial, Informática y Aeroespacial
Campus de Vegazana, s/n
24071 León}
\affiliation{IUFFyM, Universidad de Salamanca, E-37008 Salamanca, Spain}

\date{\today}

\begin{abstract}
We investigate whether a non-minimal coupling between dark matter and gravity can influence the internal dynamics of ultra-diffuse galaxies. Within this framework, the gravitational potential is modified by an additional term that captures the interaction between spacetime curvature and the dark matter with a coupling constant determined by a length scale $L$. Using spherical Jeans modelling, we analyze the kinematic data of three ultra-diffuse galaxies: NGC 1052-DF2, NGC 1052-DF4, and Dragonfly 44, which span the observational extremes from dark matter deficient to dark matter dominated systems. For each galaxy we explore several dark matter halo profiles, two orbital anisotropy models, and both with and without Stellar to Halo Mass Relation scenarios, and we perform a Bayesian parameter inference. We further validate the analysis through a sensitivity test on mock Dragonfly 44 data, which shows that only large couplings, which are already disfavored by the data, produce a detectable imprint, while smaller values remain indistinguishable from General Relativity at the current observational precision.

Across all the considered configurations, the constrained astrophysical parameters are consistent with standard ones from General Relativity. The posterior distributions of $L$ show no preference for non-zero values and result only in upper limits.
These upper limits should be interpreted as a sensitivity limit of current UDG kinematics rather than as a tight exclusion of the coupling. Future high precision velocity measurements will be essential to determine whether non-minimal coupling effects can become observationally distinguishable in low-acceleration systems.
\end{abstract}

\maketitle

\pagebreak

\section{Introduction}
\label{sec: Intro}

Astronomers have classified galaxies into a wide variety of morphological types, spanning a vast range of sizes, luminosities, and shapes. Among them low surface brightness (LSB) galaxies have long been recognized as an important part of the overall galaxy population. 

LSBs are galaxies whose central surface brightness is fainter than that of the night sky \cite{Bothun1997:LSB}. Ultra-diffuse galaxies (UDGs) represent a subclass of LSB systems, characterized by their exceptionally low central surface brightness, $\mu_{g,0} \gtrsim 24\ {\rm mag\ arcsec^{-2}}$, and relatively large effective radii, $R_{\rm e} \gtrsim 1.5\,{\rm kpc}$. However, this classification has been the subject of debate due to its somewhat arbitrary nature and the numerous observational biases influencing such selection criteria \cite{buzzo:2025multiple}, which has been covered extensively in \cite{van2022s}.

Although diffuse, extended galaxies were reported as early as the 1970s \cite{Disney1976,Bothun:1987}, they gained wide attention only after the Dragonfly Telephoto Array\footnote{\href{https://www.dragonflytelescope.org/publications.html}{https://www.dragonflytelescope.org/publications.html}} revealed an unexpectedly large population of extremely extended, faint galaxies in the Coma cluster \cite{vanDokkum:2014cea}. Initially, UDGs were sometimes dismissed as observational artifacts, but over time it became clear that they are genuine systems with distinctive properties that make them valuable laboratories for studying galaxy formation and the role of dark matter \cite{benavides2023:origin}.

Despite the progress, the nature of UDGs is poorly understood and their formation history is still uncertain. Their structural, stellar-population, and kinematic properties span a remarkably broad range, suggesting multiple formation channels. The proposed scenarios include internal processes, such as stellar feedback–driven outflows that inflate dwarf galaxies into diffuse systems (``puffed-up dwarfs”) \cite{DiCintio:2016ehs}, or the formation of extended stellar disks in high-spin DM halos \cite{amorisco:2016UDG}. Environmental (external) mechanisms such as tidal \cite{Sales:2019iwl}, ram-pressure or gas stripping by nearby massive hosts \cite{Fujita:2003zj}, and repeated high-speed encounters (``galaxy harassment”) \cite{Abraham:2014lfa,Bhatia:2024xat} have also been shown to transform compact dwarfs into diffuse systems. However, none of these mechanisms alone can account for the observed UDG diversity. As an example, low surface brightness features are widespread in the NGC 1052 group \cite{Muller:2019p}, which does not support tidal interaction as a dominant mechanism, and the dynamical data available are far too sparse to reach firm conclusions \cite{Silk:2019jbt}, hinting us that maybe a combination of formation pathways is required. Beyond their uncertain origins, UDGs are found across a wide variety of cosmic environments; from the cosmic voids \cite{Roman:2019voidUDG}, to galaxy groups \cite{Merritt:2016,SmithCastelli:2016}, dense clusters \cite{vanderBurg:2016}, and even along large scale filaments of the cosmic web \cite{MartnezDelgado:2016}.

A further complication comes from reports that some UDGs may contain very little or no dark matter. First, the evidence relies on kinematic tracers (the motions of stars or globular clusters), but these data are uncertain \cite{Laporte:2018a}. Second, the distances to some UDGs are not firmly established \cite{Trujillo:2019,nusser:2019}. If the distance is wrong, then both the assumed size and mass of the galaxy can change dramatically. Later studies have shown that neither the kinematic arguments \cite{Danieli:2019zyi} nor the distance estimates \cite{vanDokkum:2019} provide a definitive resolution.

UDGs typically lack star-forming gas \cite{vanDokkum:2014cea} or display extremely low star formation efficiency \cite{Leisman2017}. As a result, and as was mentioned earlier, they challenge our understanding of galaxy formation and evolution. Another interesting point about these objects is that they are very rich globular cluster systems that offer an opportunity to reconstruct the internal velocity dispersion of such faint galaxies \cite{Amorisco:2016sbk, Beasley:2016,PengLim:2016}. Moreover, in \cite{Bhatia:2024xat} it was shown that even for UDGs thought to be dark matter dominated, modified gravity models can fit their observed kinematics.

The idea that gravity, or the way matter couples to curvature, might deviate from the standard minimal coupling prescription of general relativity (GR) has been explored in many theoretical frameworks. Modified Newtonian Dynamics (MOND), introduced by Milgrom in 1983, proposed that galactic dynamical discrepancies may arise not from unseen dark matter but from a breakdown of Newtonian dynamics below a characteristic acceleration scale $a_0 \simeq 1.2\times10^{-10}\;\mathrm{m\;s^{-2}}$ \cite{Milgrom:1983ca,Milgrom:1983pn,Milgrom:1983zz,Famaey:2011kh}, which allows the rotation curves of galaxies and other low-acceleration systems to be reproduced without including large amounts of non-baryonic  matter. This empirical success led to several relativistic extensions (see \cite{Bekenstein:2004ne,Milgrom:2009gv}); however, none of these different types of MOND models has been able to match cosmological observations in full, especially the detailed CMB anisotropy spectrum and the matter power spectrum. Although in \cite{Skordis:2020eui} the authors propose a relativistic MOND theory that claims to be compatible with the CMB and matter-power spectrum. However, the model does not yet fully address all cosmological level problems.

An interesting viewpoint between particle DM and modified gravity consists in changing the way DM couples with curvature. One such framework is the non-minimal coupling (NMC) model introduced by \cite{Bruneton:2008fk}, where one can have the successes of the MOND at small scale and reach the $\Lambda$CDM at large scales. In this scenario, dark matter is treated as a fluid whose Lagrangian includes direct couplings to curvature scalars or tensors. These terms modify the effective gravitational potential in the weak field limit, introducing a density gradient dependent correction weighted by a characteristic length scale $L$. This modification does not add new propagating degrees of freedom and can be interpreted as a coarse-grained or fluid-averaged effect of microphysical interactions in the DM sector. If this coupling is active at galactic scales, NMC terms could change the internal kinematics of systems with shallow density profiles or low accelerations, precisely the regime where UDGs lie. 

The approach we have chosen is based on the work developed in \cite{Bettoni:2011fs,Bettoni:2015wla}. This model is particularly interesting because it is not restricted to a specific theory: similar couplings can arise in a variety of different scenarios such as those studied in \cite{Bettoni:2013zma,Ivanov:2019iec,BeltranJimenez:2017doy}. 
In this work we aim to test a non-minimal coupling model between dark matter and gravity on the scale of galaxies (see \cite{Gandolfi:2022puw} for an analysis of NMC in spiral galaxies).  We have chosen to start with the UDGs precisely because of the extreme variety in their kinematical and environmental properties (apart from photometric classification), and because of the large variation in dark matter content within them. This makes them the perfect benchmark for testing gravity.

The paper is structured as follows: First, in Sec.\;\ref{sec: Theo} we briefly review the theory in use. In Sec.\;\ref{sec: GD}, we describe the kinematic properties of UDGs analyzed in this work and the modelling of their internal structure. Sec.\;\ref{sec: Data} summarizes the main characteristics of the three UDGs studied and introduces the datasets used in our analysis. The key elements of our statistical framework are presented in Sec.\;\ref{sec: SA}. The results are discussed in Sec.\;\ref{sec: Discu} and our main conclusions are in Sec.\;\ref{sec: Conc}.

\section{Theoretical Background}
\label{sec: Theo}

We explore an extension of GR where DM interacts non-minimally with spacetime curvature. The general action for this NMC model is given by \cite{Bettoni:2015wla} 
\be
\label{eq: action_all}
S = S_{\rm EH} + S_{d} + S_{c} + S_{m}\, ,
\ee
where $S_{\rm EH}$ is the Einstein--Hilbert action, and $S_m$ describes the DM as a perfect fluid \cite{Brown:1992kc}. The additional terms represent the non-minimal couplings:
\begin{itemize}
    \item $S_d$ introduces a disformal coupling that links the DM four-velocity ($u^\mu$) and density to the Ricci tensor  $R_{\mu\nu}$ \footnote{See \cite{Bekenstein:1992pj} for a  thorough introduction to disformal transformations.} (see \cite{Bettoni:2011fs,Bettoni:2015wla} for a detailed discussion):
    \be
    S_{d} = \frac{M_{\textrm{Pl}}^2}{2}\int d^4 x\sqrt{-g} \big[\alpha_{d} F_d (\rho) R_{\mu\nu}u^\mu u^\nu \big]\, ,
    \ee
    \item $S_c$ introduces a conformal coupling that links DM density to the Ricci scalar $R$:
    \be
    S_{c} = \frac{M_{\textrm{Pl}}^2}{2}\int d^4 x\sqrt{-g} \big[\alpha_{\mathrm{c}} F_c (\rho) R\big]\, .
    \ee
\end{itemize}
The parameters $\alpha_d$ and $\alpha_c$ measure the coupling strength and the functions $F_d(\rho)$ and $F_c(\rho)$ can be arbitrary functions of fluid variables but for simplicity, they are set proportional to the DM density, $F_i (\rho)\propto \rho_{{\rm DM}}$\footnote{For more information look at \cite{Bettoni:2011fs}} \cite{Bettoni:2015wla}. This assumption ensures that the couplings only affect dark matter, leaving baryons to behave as in the standard model.

\section*{Newtonian Limit}

To understand the impact of these couplings at galactic scales, we examine the Newtonian limit, described by the metric potentials $\Phi$ and $\Psi$ in the Newton gauge
\be
ds^2 = -(1+2 \Psi)dt^2+(1-2 \Phi)\delta_{ij}dx^idx^j\, .
\ee
The NMCs lead to modifications to the gravitational equations \cite{Bettoni:2011fs,Bettoni:2015wla} as it is summarized below.

\paragraph*{\textbf{a. Disformal Coupling:}}
A pure disformal coupling modifies the Poisson equation as
\be
\label{eq: poissonmoddisf}
\mathbf{\nabla}^2 \Phi = 4\pi G\,\Big[\rho_{\mathrm{tot}} - \epsilon\, L^2\, \nabla^2 \rho_{\rm DM}(r) \Big]\, ,
\ee
where $\rho_{\mathrm{tot}} = \rho_{\rm bar} + \rho_{\rm DM}$ and $L$ is a characteristic length scale that relates with the original coupling constant $\alpha_d$ via the relation $\alpha_d F_d = -8\pi G L^2\rho_{\rm DM}$. This equation shows that the gravitational field is sourced not only by the total mass but also by spatial inhomogeneities in the DM distribution ($\nabla^2 \rho_{\rm DM}$). This effect is stronger in regions with lumpy DM density.

One of the key features of this model is the absence of anisotropic stress, meaning that $\Phi = \Psi$. Consequently, the Weyl potential $(\Phi+\Psi)/2$, which is probed by gravitational lensing, is also modified by the DM inhomogeneities
\be
\label{eq: Weyl_disf}
\nabla^{2} \left( \frac{\mathrm{\Phi}+\mathrm{\Psi}}{2} \right) = 4\pi G\,\Big[\rho_{\rm tot} - \epsilon\,L^2\, \nabla^2 \rho_{\rm DM}(r) \Big]\, .
\ee

\paragraph*{\textbf{b. Conformal Coupling:}}
In the case of a conformal coupling, the modified Poisson equation takes the same functional form as in the disformal scenario. However, this coupling introduces anisotropic stress ($\Phi \neq \Psi$), with the potentials related by $\Psi = \Phi - \alpha_{c} F_c$ where we have again expressed the original coupling as $\alpha_c F_c = -8\pi G L^2\rho_{\rm DM}$. As a result, when calculating the Weyl potential, the NMC terms cancel, and the equation becomes identical to that of GR. Thus, a conformal coupling has no observable effect on gravitational lensing \cite{Bettoni:2011fs}.

\paragraph*{\textbf{c. Einstein Tensor Coupling:}}
A third scenario involves a coupling to the Einstein tensor ($G_{\mu\nu}$), which can be achieved by setting $\alpha_c = -\alpha_d/2$ and $F_{c} = F_{d} = F_{E}$ \cite{Bettoni:2011fs,Bettoni:2015wla}. The gravitational part of the action then reads
\be
\label{eq: action_con}
        S = \frac{M_{\textrm{Pl}}^2}{2}\int d^4 x\sqrt{-g} \,\left[ R +\alpha_{E} F_{E}(\rho) G_{\mu\nu} u^\mu u^\nu \right]
        \; .
\ee

This modification leaves the Poisson equation unchanged from GR, but the presence of anisotropic stress leads to a modified Weyl potential similar to the disformal case. Therefore, gravitational lensing observations cannot distinguish between a disformal coupling and an Einstein tensor coupling.

The parameter $L$ is a characteristic length scale which could be a local property of the environment. In contexts like Bose--Einstein condensates, $L$ is analogous to a ``healing length" and is related to the microscopic properties of the underlying particles \cite{Bettoni:2013zma,Boehmer:2007um}.

\subsection{Choice of coupling}

The three coupling scenarios summarized above lead to different modifications of the metric potentials and therefore to different observational signatures. Since the present work aims at testing NMC through its impact on internal kinematics of stars and globular clusters, the relevant quantity is the dynamical potential entering the Jeans equation.

For this reason, in what follows we focus on the disformal coupling only. This choice is motivated by the fact that it is the scenario whose modifications appear in both the dynamical and the lensing sectors, making it directly comparable with our previous cluster-scale analyses \cite{Zamani:2024oep,Zamani:2024qbx}. However, since the three couplings corrections differ only by an $\mathcal{O}(1)$ coefficient, our disformal bounds on $L$ translate to the other cases through a simple rescaling, so the constraints reported here effectively cover all three coupling scenarios.
\section{Galaxy Dynamics}
\label{sec: GD}

To investigate the internal kinematics of our sample of ultra--diffuse galaxies (UDGs), we need to solve the Jeans equation. Assuming spherical symmetry and the absence of net streaming motions \cite{Mamon:2004xk}, the Jeans equation is
\be
\frac{d}{dr} \left[\ell(r) \sigma_r^2(r)\right] + \frac{2\beta(r)}{r} \ell(r) \sigma_r^2(r) = -\ell(r) \frac{d\Phi(r)}{dr},
\ee
where $\ell(r)$ is the luminosity density, $\sigma_r(r)$ is the radial velocity dispersion, $\Phi(r)$ is the gravitational potential, and $\beta(r)$ is the anisotropy parameter defined as
\be
\beta(r) = 1 - \frac{\sigma_t^2(r)}{\sigma_r^2(r)},
\ee
with $\sigma_t(r)$ showing the tangential velocity dispersion.

When we observe galaxies, the main quantity we can measure is the line of sight velocity dispersion $\sigma_{\text{los}}(R)$. However, this observable depends not only on the galaxy’s mass distribution $M(r)$ but also on the orbital structure of its stars, described by the anisotropy profile $\beta(r)$. This creates what is known as the mass-anisotropy degeneracy (see \cite{Michael:Degeneracy1990,Lokas:2003ks,Wilkinson2002:DMdSphI}): a galaxy with a large mass but predominantly tangential orbits may yield the same $\sigma_{\text{los}}(R)$ as a galaxy with a smaller mass but mainly radial orbits. The reason for that is that radial orbits contribute more strongly to the velocity dispersion along the line of sight, while tangential orbits tend to suppress it. At the same time, changing the mass distribution modifies the gravitational potential, which changes the underlying radial dispersion $\sigma_r(r)$. Because these effects can compensate for one another, different combinations of $M(r)$ and $\beta(r)$ can reproduce very similar observational signatures, making it difficult to disentangle mass from orbital structure without additional constraints.

Several methods beyond the spherical Jeans equation have been developed to constrain mass profiles from velocity dispersions including virial estimators \cite{Illingworth1976,Heisler:1985} distribution-function methods \cite{Agnello:2012uc,Amorisco:2011hb}, Schwarzschild orbit-superposition modelling \cite{Breddels:2013dga, Breddels:2012cq}, and higher-order velocity-moment methods \cite{Merrifield:1990,Richardson:2012ig,Richardson:2014mra}. However, given the limited number of kinematic tracers available (10 GCs for NGC 1052-DF2, 7 for NGC 1052-DF4, and 9 radial bins for Dragonfly 44) the spherical Jeans equation represents one of the most  statistically appropriate frameworks for our datasets \cite{Laporte2019}. A known limitation of our approach is the mass--anisotropy degeneracy \cite{Strigari:2007vn, Watkins:2013fna}. Methods such as higher-order velocity moments and axisymmetric Jeans modelling \cite{Read:2017} can help resolve this degeneracy when sufficiently rich datasets are available. We further reduce the mass--anisotropy degeneracy by adopting two different anisotropy parametrizations and physically motivated priors on the halo mass and concentration.

Due to the degeneracy between these functions, we consider two commonly adopted parametrization for $\beta(r)$ 
\be
\beta(r) = \begin{cases}
\beta_c, & \text{constant anisotropy} \\
\beta_0 + (\beta_\infty - \beta_0) \; \frac{r}{r + r_a}, & \text{radial profile}
\end{cases}
\ee
where $r_a$ is the anisotropy radius controlling the transition rate between the inner ($\beta_0$) and outer ($\beta_\infty$) asymptotic anisotropy values. The radial anisotropy profile was first presented in \cite{Napolitano:2008ft} and later applied successfully to UDGs in \cite{Zhang:2015pca}.
 
In General Relativity, having $\frac{d\Phi}{dr} = \frac{G M(r)}{r^2}$, the general solution for the Jeans equation to get the radial velocity dispersion is \cite{Mamon:2012yb}
\be
\ell(r) \sigma_r^2(r) = \frac{1}{f(r)} \int_r^\infty ds\, f(s) \ell(s) \frac{G M(s)}{s^2},
\ee
where the integrating factor $f(r)$ is defined by the anisotropy profile through the relation
\be
\frac{d \ln f(r)}{d \ln r} = 2 \beta(r).
\ee

Since observations yield the line of sight velocity dispersion $\sigma_{\rm los}(R)$, we project the 3D velocity field along the line of sight 
\begin{align}
\label{eq: sigmalos1}
\sigma_{\rm los}^2(R)
&= \frac{2}{I(R)} \int_R^\infty dr \,
\Bigg[
\frac{r\,\ell(r)\sigma_r^2(r)}{\sqrt{r^2-R^2}}\\ \nonumber
& \qquad \qquad \qquad \qquad 
- \frac{R^2\,\beta(r)\,\ell(r)\sigma_r^2(r)}{r\sqrt{r^2-R^2}}
\Bigg].
\end{align}
where $R$ is the projected radius and $I(R)$ is the surface brightness profile.

We can rewrite Eq. (\ref{eq: sigmalos1}) in a more compact expression that can be obtained by defining a kernel function $K\left(\frac{r}{R}, \frac{r_a}{R}  \right)$ \cite{Mamon:2004xk}
\be
\label{eq: sigmalos2}
\sigma_{\rm los}^2(R) = \frac{2G}{I(R)} \int_R^\infty dr \, K\left(\frac{r}{R}, \frac{r_a}{R}\right) \frac{\ell(r) M(r)}{r}.
\ee

In theories beyond General Relativity the gravitational potential $\Phi(r)$ no longer obeys the standard Poisson equation. Instead, one can introduce an effective mass profile $M_{\rm eff}(r)$ that can absorb the correction of NMC into itself as 
\be
\label{eq:Meff1}
\frac{d\Phi}{dr} = \frac{G M_{\rm eff}(r)}{r^2}\, ,
\ee
so that the impact of the modifications is fully captured by the effective mass profile $M_{\rm eff}(r)$, while the standard Jeans formalism can be applied without further modification. In our case we have
\be
\label{eq:Meff2}
M_{\rm eff}(r) = M(r) - 4\pi \epsilon r^2 L^2 \frac{d \rho_{\rm DM}}{dr}\,.
\ee

Substituting this into the projected Jeans expression gives
\be
\sigma_{\rm los}^2(R) = \frac{2 G}{I(R)} \int_R^\infty dr \, K\left(\frac{r}{R}, \frac{r_a}{R}\right) \frac{\ell(r) M_{\rm eff}(r)}{r},
\ee
providing a pathway to test alternative gravity theories through observable kinematic profiles of UDGs.

\subsection{DM halo density Profile}
\label{sec: NFW}

In this work, we consider eight different types of DM profiles. Among them, the generalized Navarro-Frenk-White (gNFW) profile is the most commonly used profile in the literature on galactic scales. As a sample of those eight profiles, we presented the results of gNFW and NFW in detail. The information regarding the rest of the profiles is presented in the \ref{app: DMHalo}. 

The NFW profile \cite{Navarro:1995iw,Navarro:1996gj} is described as 
\begin{equation}
    \rho_{\mathrm{NFW}}(r)  = \frac{\rho_s}{\dfrac{r}{r_s}\left(1 + \dfrac{r}{r_s}\right)^2},
\end{equation}
with
\begin{align}
    \rho_s &= \frac{\Delta}{3}\,\rho_c \, \frac{c^3}{\ln(1 + c_\Delta) - \dfrac{c_\Delta}{1 + c_\Delta}}, \nonumber \\[3ex]
    M(<r) & = 4\pi \rho_s r_s^3 \left[\ln\!\left(1+\frac{r}{r_s}\right)-\frac{r/r_s}{1+r/r_s}\right]\, ,
\end{align}
while the generalized NFW (gNFW) profile \cite{Umetsu:2015baa,Zhao:1995cp} is defined by
\be
\rho(r)=\frac{\rho_s}{\big(\frac{r}{r_s}\big)^{\gamma} \big(1+\frac{r}{r_s}\big)^{3-\gamma}}\;,
\ee
with
\begin{align}
    \rho_s &=\frac{{\rm \Delta}}{3}\rho_{c} c_{{\rm \Delta}}^{\gamma} \frac{(3-\gamma) (2-\gamma)^{\gamma}}{ 
 _{2}F_{1}[3-\gamma,3-\gamma,4-\gamma,(\gamma-2) c_{{\rm \Delta}}]}\, \nonumber \\[3ex]
    M(<r)&= 4\pi \rho_s r_s^3 \left( \frac{r}{r_s} \right)^{3 - \gamma}\, \nonumber \\[-1.5ex]
    & \qquad \qquad \qquad {}_2F_1\!\left( 3 - \gamma,\, 3 -\gamma;\, 4 - \gamma;\, -\frac{r}{r_s} \right),
\end{align}
where $_{2}F_{1}$ is the Gaussian hypergeometric function, and $0<\gamma<2$. 

The radius $r_s$ is the scale parameter that determines how rapidly the halo density changes from the inner to the outer regions, but it can have a different meaning depending on the chosen model. To conduct a consistent comparison of various DM profiles, we normalize all models to the same quantity, known as $r_{-2}$, which has a more direct physical meaning (\textit{e.g.} see \cite{Ascasibar:2003mm}): it shows the specific radius where the logarithmic slope of the density profile is equal to $-2$, which means 
\be
\label{eq: slope}
\left. \frac{d\ln \rho(r)}{d\ln r} \right|_{r = r_{-2}} = -2 .
\ee

In some models, such as the standard NFW profile, the slope equals $-2$ precisely at $r = r_s$, so the two radii are the same. However, in some profiles where the inner slope or the transition between inner and outer regions is different, the slope reaches $-2$ at a different radius, leading to $r_{-2} \neq r_s$. For the gNFW, for example, we have $r_{-2} = (2-\gamma) r_s$.

\subsection{Stellar Mass}
We also consider the stellar contribution in the Jeans equation. Starting from the S\'{e}rsic surface brightness Profile \cite{sersic:1963a,sersic:1968b}

\be
\label{eq: sersic}
    I(R) = I_0 \exp \left[ - \left(\frac{R}{a_s}\right)^{1/n} \right] \, ,
\ee
which $a_s$ is the S\'{e}rsic scale parameter, $R$ is projected (2D) radial distance, $n$ is S\'{e}rsic index and $I_0$ is the central surface brightness of the galaxy, defined as
\be
    I_0 = \frac{L_{\text{tot}}}{2 \pi \, n \, a_s^2 \, \Gamma[2 n]} \, ,
\ee
and the $L_{\text{tot}}$ is the total luminosity of the galaxy
\be
    L_{\text{tot}} = 10^{-0.4 [m_{V_{606}} - \mu(\mathcal{D}) - M_{\odot , V_{606}}]} \,,
\ee
Here, $m_{V_{606}}$ is the apparent magnitude of the galaxy, and the distance modulus is given by $\mu(\mathcal{D})= 5 \log_{10}(\mathcal{D})+25$, where $\mathcal{D}$ is the galaxy's distance in Mpc. $ M_{\odot, V_{606}} = 4.72$ shows the absolute magnitude of the Sun in the V-band (HST/ACS F606W filter) \cite{Willmer:2018a}.
In addition we can write $a_s$ as
\be
a_s=\frac{R_{e}}{(b_n)^n}
\ee
with $R_e$ being the effective radius that encloses half of the galaxy's total luminosity, and $b_n$ is a function of the shape parameter $n$, such that $\Gamma(2n) \;=\; 2\,\gamma\!\left(2n,\, b_{n}\right)$ where $\Gamma$ is the gamma function and $\gamma$ is the incomplete gamma function that can be approximated as $b_n=2n-0.33$ \cite{Caon:1993wb}.
Now, in order to get the stellar luminosity density profile we need to do the deprojection from 2D to 3D using the Abel transform of Eq. (\ref{eq: sersic}). Although it is not possible to derive an analytical expression for the S\'ersic index, we can approximate it numerically with
\be
\ell_{\ast}(r)= \ell_1 \tilde{\ell}\left(\frac{r}{a_s}\right)
\ee
with
\begin{align}
\tilde{\ell}(x) &= x^{-p_n} \, \exp(-x^{1/n}), \\
\ell_{1} &= \frac{L_{\text{tot}}}{4 \pi n \, \Gamma\!\left[(3 - p_n)n\right] \, a_{s}^{3}} .
\end{align}
and the function $p_n$ is defined in Eq. (19) of \cite{Neto:1999gx} as
\be
p_n = 1.0 - \frac{0.6097}{n} + \frac{0.05463}{n^2}.
\ee

Finally, in order to have the mass density, we need to multiply by the mass--to--light ratio ($\Upsilon_{\ast}$) 
\be
\rho_{\ast}(r)= \Upsilon_{\ast} \, \ell_1 \tilde{\ell}\left(\frac{r}{a_s}\right)
\ee

{\scriptsize
\setlength{\tabcolsep}{2.0mm}
\renewcommand{\arraystretch}{2.25}
\begin{table}[t]
\centering
\begin{threeparttable}
\caption{General properties of DF2 \cite{vanDokkum:2018vup}, DF4 \cite{vanDokkum:2019}, and DF44 (V606 band).}
\label{tab:UDG_charac}
\vspace{5pt}
\begin{tabular}{c|ccc}
\tightdoublehline
Parameter & DF2 & DF4 & DF44 \\
\tightdoublehline
$z$ & 0.005\tnote{a} & 0.005 & 0.023156\tnote{b} \\
\hline
Sérsic index $n$ & 0.6 & 0.79 & 0.94 \\
\hline
$R_\mathrm{eff}$ (arcsec) & 22.6 & $\sim$16.58 & $\sim$9.69 \\
\hline
$R_\mathrm{eff}$ (kpc) & --- & 1.6 & 4.7 \\
\hline
$D$ (Mpc) & 20 & 20 & 100 \\
\hline
Axis ratio $b/a$ & 0.85 & 0.89 & 0.68 \\
\hline
$\mu_0$ (mag arcsec$^{-2}$) & 24.4 & 23.7 & 24.1 \\
\hline
$M_V$ & $-15.4$ & $-15.0$ & $-16.2$ \\
\hline
$L_\mathrm{tot}$ ($L_\odot$) & $1.12\times10^8$ & $7.7\times10^7$ & $2.33\times10^8$ \\
\hline
\end{tabular}

\begin{tablenotes}[flushleft]
\footnotesize
\item[a] Taken from \href{https://ned.ipac.caltech.edu/byname?objname=NGC+1052&hconst=73&omegam=0.27&omegav=0.73&wmap=1&corr_z=1}{NED database}.
\item[b] Taken from SIMBAD catalogue \cite{Wenger:2000sw}.
\end{tablenotes}

\end{threeparttable}
\end{table}}

\section{Data}
\label{sec: Data}

In this study, we consider three different UDGs: NGC 1052-DF2, NGC 1052-DF4 ($\lesssim 1 \% $ \cite{Montes:2020zaz}), which have been considered to lack dark matter, and Dragonfly 44 ($\sim 98\%$ DM content \cite{vanDokkum:2016uwg}), which is thought to be dark matter dominated. While the detailed properties of each UDG are presented in Table \ref{tab:UDG_charac}, here we provide a brief overview of how the data were collected and the main studies that reported them.

NGC 1052-DF2 was first observed by the Dragonfly Telephoto Array \cite{Abraham:2014lfa}. The structural properties of DF2 were further studied using Hubble Space Telescope photometry \cite{vanDokkum:2018vup}. In addition, spectroscopy of NGC 1052-DF2's globular cluster was conducted with the W. M. Keck Observatory's twin telescopes \cite{vanDokkum:2018vup}. Ten globular clusters with measured radial velocities were used in the original kinematic analysis of NGC 1052–DF2, as reported in \cite{vanDokkum:2018vup}\footnote{Although the velocity dispersion values are not explicitly stated in the paper, they are shown at the top of Figure 6 on the author's website: \href{https://www.pietervandokkum.com/ngc1052-df2}{www.pietervandokkum.com/ngc1052-df2}. Note that one of the velocity measurements was later revised, as discussed in \cite{vanDokkum:2018DF2}.}. Although imaging data suggest the presence of more globular clusters in NGC 1052–DF2, as mentioned earlier, only ten were spectroscopically confirmed and used in the original velocity dispersion analysis by van Dokkum \cite{vanDokkum:2018vup}. A recent study analyzing archival MUSE data identified more clusters with confirmed radial velocities, increasing the total number of spectroscopically confirmed globular clusters to 16 \cite{Fahrion2025:DF2}. The kinematic properties of DF2 have been investigated in  \cite{Wasserman:2018scp}. Furthermore, both DF2 and DF4, have been jointly analyzed in \cite{Forbes:2019fxt}. A more recent dynamical study of DF2 using anisotropic distribution functions \cite{Aditya:2024wdw} found tangentially biased stellar orbits and discussed the degeneracy between cored and cuspy halo models.

NGC 1052-DF4 is a diffuse galaxy associated with the NGC 1052 group  (same as NGC 1052-DF2). Deep spectroscopic observations carried out with the \textit{Keck\;I Telescope} using the Low Resolution Imaging Spectrograph (LRIS) have revealed that this system hosts a small population of massive and compact globular clusters, extending up to nearly $7\;\mathrm{kpc}$ from the galaxy's center \cite{vanDokkum:2019}. These findings provide valuable evidence for its unusually low dark matter content. The data that was used for our statistical analysis are taken from Table 1 of \cite{vanDokkum:2019}.

Dragonfly 44: the spectroscopic observations were obtained using the \textit{Keck Cosmic Web Imager (KCWI)} using the Keck\;II Telescope \cite{vanDokkum:2019fdc}. This UDG is located within the Coma cluster. For our analysis, we employed the kinematic measurements listed in Table\;2 of \cite{vanDokkum:2019fdc}, based on spatially resolved integrated stellar light rather than discrete globular cluster tracers, which provide the radial velocity and velocity dispersion profiles across nine distinct radial intervals.

\section{Statistical Analysis}
\label{sec: SA}

{\renewcommand{\tabcolsep}{1.5mm}
{\renewcommand{\arraystretch}{2.}
\begin{table}[h!]
\begin{minipage}{0.45\textwidth}
\centering
\caption{Prior distributions used throughout our statistical analysis. 
$\mathcal{N}$ denotes normal priors, 
$\log\mathcal{N}$ log-normal priors (Gaussian in the log-variable), 
and $\mathcal{U}$ uniform priors. 
Positivity is enforced for $v_{\rm sys}$, $D$, and $\Upsilon_{\ast}$. 
Here $f_c(M_{200})$ denotes the $c$--$M$ relation from \cite{Correa:2015dva}, 
and $f_M(D,\Upsilon_{\ast})$ the stellar to halo mass relation (SHMR) from \cite{Rodrguez:2017}.}
\label{tab: priors}
\resizebox*{\textwidth}{!}{
\begin{tabular}{ccc}
\tightdoublehline
Parameter & Prior & Reference \\
\tightdoublehline
\multicolumn{3}{c}{Global priors (all galaxies)} \\
\tightdoublehline
$c_{200}$ & $\log\mathcal{N}(f_c(M_{200}),\,0.16) \in (0,\,40) $ &  \cite{Correa:2015dva} \\

$M_{200}$ & $\mathcal{U}(0,\,20)$ & $-$ \\

$\gamma$ (gNFW) & $\mathcal{U}(0,\,2)$ & $-$ \\

$\beta_i$ & $\log\mathcal{N}(0,\,0.5) \in (-10,\,1)$ & \cite{Wasserman:2018scp} \\

$r_a$ (kpc) & $\mathcal{U}(0,\,\infty)$ & $-$ \\

$\log M_\ast$ & $\mathcal{N}(f_M(D,\Upsilon_\ast),\,0.3)$ & \cite{Rodrguez:2017} \\
\tightdoublehline
\multicolumn{3}{c}{Galaxy-specific priors} \\
\tightdoublehline
\multirow{3}{*}{$D$ (Mpc)} 
   & DF2: $\mathcal{N}(22.1,\,1.2)$ & \cite{shen2021:tip} \\
   & DF4: $\mathcal{N}(22.1,\,1.2)$ & \cite{shen2021:tip} \\
   & DF44: $\mathcal{N}(102,\,14)$  & \cite{Thomsen:1997vz} \\
\hline
\multirow{2}{*}{$v_{\rm sys}$ (km s$^{-1}$)} 
   & DF2: $\mathcal{N}(1801.6,\,5.0)$ & \cite{Wasserman:2018scp} \\
   & DF4: $\mathcal{N}(1444.60,\,7.75)$ & \cite{vanDokkum:2019} \\
\hline
\multirow{3}{*}{$\Upsilon_\ast$}
   & DF2: $\mathcal{N}(1.7,\,0.5)$ &  \cite{Wasserman:2018scp} \\
   & DF4: $\mathcal{N}(2.0,\,0.5)$ &  \cite{vanDokkum:2019} \\
   & DF44: $\log\mathcal{N}(1.5,\,0.1)$ & \cite{wasserman2019:spatially} \\
\hline
\end{tabular}}
\end{minipage}
\end{table}}}

Here we want to constrain the parameters of the NMC model, considering different dark matter profiles. In order to do so, for DF2 and DF4 we consider a Gaussian likelihood and the $\chi^2$ can be written as
\be
\chi^{2}(\boldsymbol{\theta}) = \sum_{i=1}^{N_{\rm data}}
\left[\frac{(v_i - v_{\rm sys})^{2}}{\sigma_i^{2}(\boldsymbol{\theta})}+ \ln\!\big(2\pi \sigma_i^{2}(\boldsymbol{\theta})\big)\right],
\label{eq:chi2}
\ee
where: $N_{\rm data} = 10$ and $7$ for DF2 and DF4, respectively, is the number of globular clusters within them; $v_{\rm sys}$ is the systemic velocity of the UDGs;
$\sigma_i^{2}(\boldsymbol{\theta}) \equiv \sigma_{ los,i}^{2}(\boldsymbol{\theta}) + \sigma_{v_i}^{2}$, is the total variance on the velocities $v_i$, with measurement errors $\sigma_{v_i}$ and measured velocity dispersions $\sigma^2_{los}(\boldsymbol{\theta})$; $\boldsymbol{\theta}$ is the vector of free parameters that we have in our model. When considering the GR case, with a gNFW--type profiles and with constant anisotropy, we have $\boldsymbol{\theta} = \{c_{200},M_{200}, \gamma,\beta_c,D,\Upsilon_\ast,v_{\rm sys}\}$, while for radial anisotropy $\boldsymbol{\theta} = \{c_{200},M_{200}, \gamma,\beta_0,\beta_{\infty},r_a,D,\Upsilon_\ast,v_{\rm sys}\}$. When using the NFW profile, of course we will not have the $\gamma$ as our DM parameter. When we consider the NMC scenario, the new parameter $L$ will be added to the previous combination. 

For DF44, we instead compare effective dispersions
\be
\chi^{2}(\boldsymbol{\theta}) = \sum_{i=1}^{N_{\rm data}}
\left[\frac{\sigma_{{\rm eff},i} - \sigma_{\rm los,i}(\boldsymbol{\theta})}{\delta\sigma_{{\rm eff},i}}\right]^{2},
\label{eq:chi44}
\ee
where $\sigma_{{\rm eff},i}^{2} \equiv \sigma_{i}^{2} + v_i^{2}$ and $\delta\sigma_{{\rm eff},i}$ is its uncertainty.

A detailed list of the priors that we have considered in our analysis can be seen in  Table\;\ref{tab: priors} together with the catalogs and references from which they are taken. We want to highlight the importance of two priors specifically. UDGs are supposed to be embedded in dark matter halos that are expected to extend well beyond the radial range probed by the data. To address this limitation and preserve the physical interpretability of the dark matter parameters, we impose a log--normal prior on the concentration parameter $c_{200}$, based on the $c$--$M$ relation of \citep{Correa:2015dva}, $f_c(M_{200})$. We adopt the version recalibrated to the \textit{Planck} 2015 cosmology (see their Appendix B1), which spans masses and redshifts more suitable for our galaxy sample than other relations available in the literature. The applied scatter is $\sigma_{\log c_{200}} = 0.16$ dex.

Moreover, we consider two approaches, whether or not we correlate the stellar mass, $M_{\ast} = \Upsilon_{\ast} L_{tot}$, and the virial mass $M_{200}$. In the first case, the two masses are linked by the stellar component via the Stellar to Halo Mass Relation (SHMR) of \citep{Rodrguez:2017}, $M_{\ast} = f_M(\Upsilon_{\ast},M_{200})$ and we adopt a log--normal prior on $M_{\ast}$ with a mean given by their relation and a scatter of $0.3$ dex. In the second approach, no correlation is assumed.

Finally, we considered a series of physically motivated controls, like forcing the line of sight velocity dispersion to remain strictly positive, $\sigma_{\mathrm{los}} > 0$, at all radii sampled by the data. 

We explored the parameter space using our own code for running Markov Chain Monte Carlo (MCMC). The chain convergence is checked using the method introduced in \citep{Dunkley:2004sv}. In order to establish a statistically reliable hierarchy of scenarios that better agree with the data, we rely on the Bayes Factor \citep{Kass:1995loi}, $\mathcal{B}^{i}_{j}$, defined as the ratio between the Bayesian Evidences of scenario $\mathcal{M}_i$ and scenario $\mathcal{M}_j$. The evidence is calculated numerically using our own code implementing the Nested Sampling algorithm developed by \citep{Mukherjee:2005wg}. The Bayes Factor is then interpreted in terms of the empirical Jeffreys' scale \citep{Jeffreys1939-JEFTOP-5}.

\subsection{Sensitivity Test}
\label{sec:Sens}

\begin{figure*}
    \centering
    \begin{subfigure}{\textwidth}
        \centering
        \includegraphics[width=0.7\textwidth]{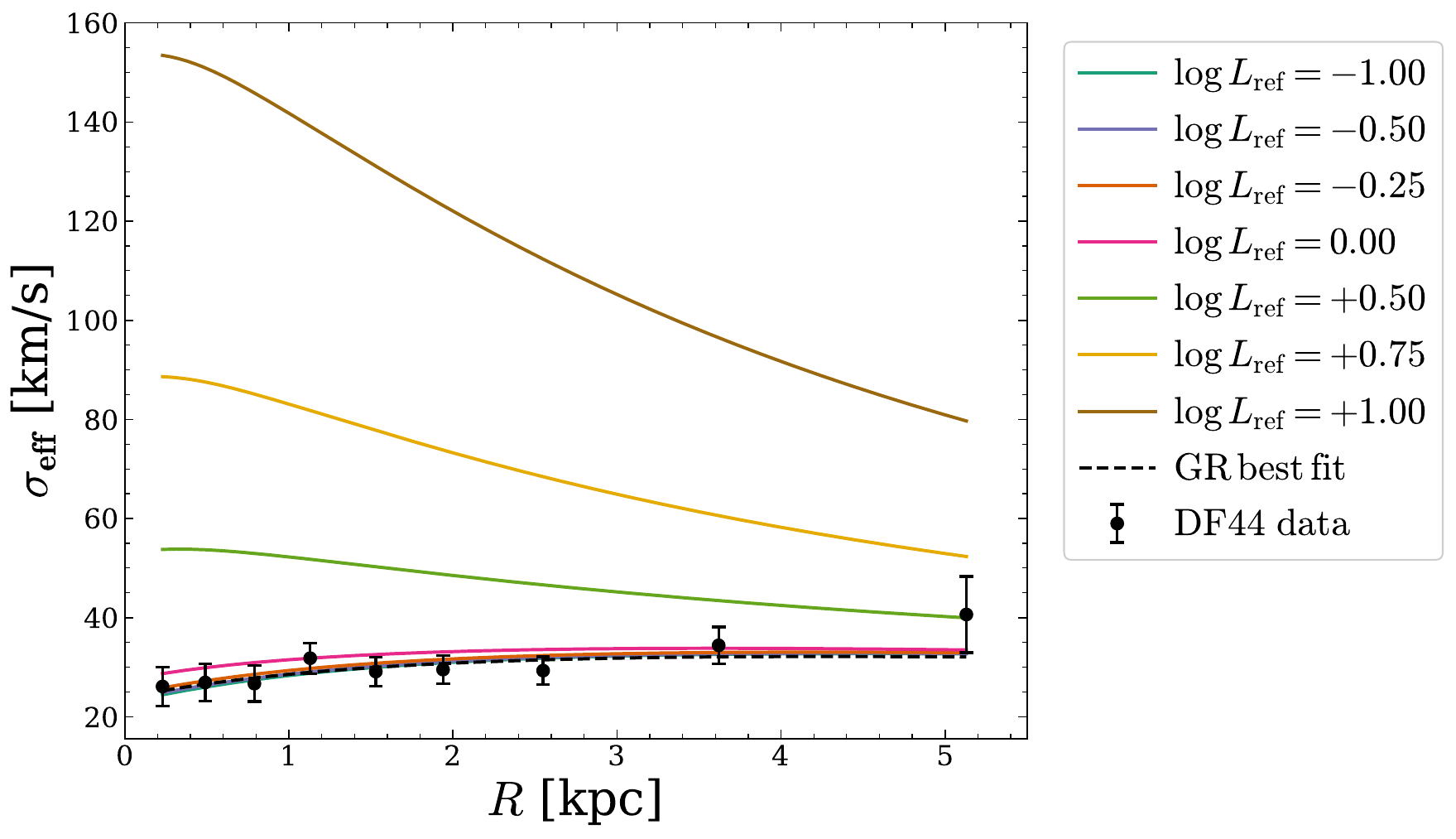}
        \caption{Predictions for the full range of referenced couplings, $\log L_{\rm ref} \in \{-1.0, -0.5, -0.25, 0.0, +0.5, +0.75, +1.0\}$, chosen to properly sample the  regime where the NMC contribution is most easily confused with GR, plus two strong-coupling cases as positive controls. The black dashed line shows the GR best-fit, and black points are the DF44 measurements from \cite{vanDokkum:2019fdc}.}
        \label{fig:sigma_all}
    \end{subfigure}
    
    \vspace{0.5cm}
    
    \begin{subfigure}{\textwidth}
        \centering
        \includegraphics[width=0.7\textwidth]{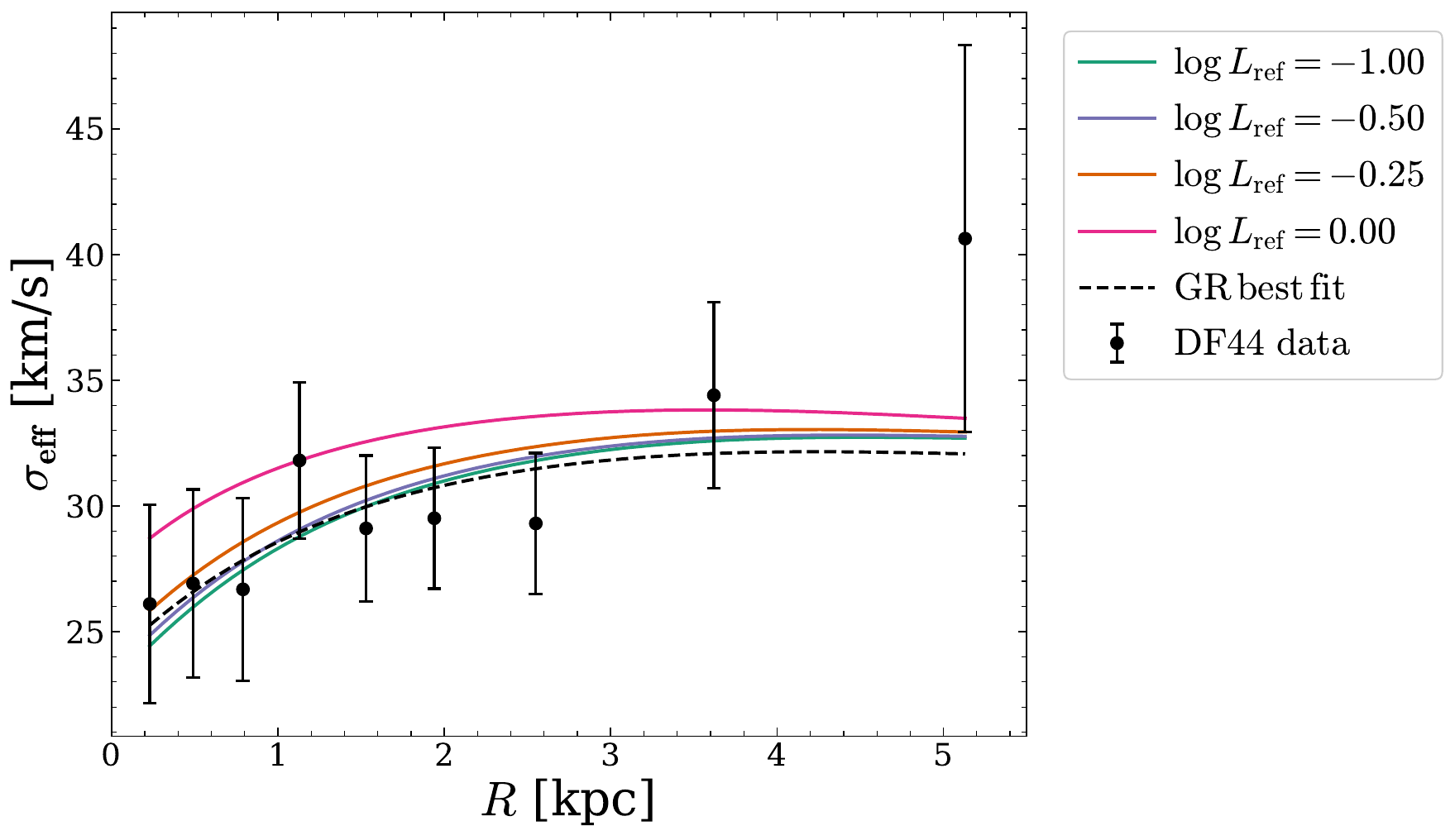}
        \caption{Zoom into the values $\log L_{\rm ref} \le 0$.}
        \label{fig:sigma_negL}
    \end{subfigure}
    
    \caption{Dragonfly 44 line of sight velocity dispersion as a function of projected radius.}
    \label{fig:df44_sigma}
\end{figure*}

In order to verify if an NMC correction can be detected by the considered data, namely to assess on more solid quantitative grounds which range of $L$ is to be waited to effectively affect the dynamical profile of the UDGs, we performed an sensitivity test on synthetic (mock) datasets. 
We chose Dragonfly 44 as the case study for this exercise because it is dark matter dominated and is therefore the case in which any potential NMC effect would/could be most prominent. The mock datasets were built to have the DF44's observational setup (same radial bins and uncertainties) so that the recovered sensitivity threshold reflects what data of this quality and sampling can detect. As reference model, we adopted the gNFW profile with the SHMR prior, constant velocity anisotropy, and a positive coupling signature ($\epsilon > 0$). 
We produced the mock data for $\sigma_{\mathrm{los}}$ for a set of reference coupling values $\log L_{\mathrm{ref}} \in \{-1.0, -0.5,-0.25, 0.0, +0.5, +0.75, +1.0\}$. For each value of $L_{\mathrm{ref}}$ we proceeded as follows:
\begin{enumerate}
    \item First, we fixed the remaining six parameters $\{c_{200}, M_{200}, \gamma, D, \beta, \Upsilon_\star\}$ to their best-fit values from the real data DF44 analysis.
    \item We computed the corresponding noise-free velocity-dispersion profile $\sigma_{\mathrm{eff}}(R_i)$ at the nine observed radial bins of DF44.
    \item We added independent Gaussian noise to each bin, drawn from the published observational uncertainty $\delta\sigma_{\mathrm{eff}}(R_i)$. This produced a synthetic dataset statistically equivalent to the real DF44 data, except that the true value of $L$ is known and controlled by us.
    \item We analyzed each mock dataset with the same MCMC pipeline used in the main analysis, leaving all seven parameters free to vary.
    \item At the end, we compared the recovered marginal posterior of $\log L$ with the referenced value $\log L_{\mathrm{ref}}$ to see the sensitivity limit.
\end{enumerate}
As shown in Fig. \ref{fig:df44_sigma}, the outcome of this test depends strongly on the magnitude of the referenced coupling. For $\log L_{\mathrm{ref}} < 0.5$ the NMC contribution becomes negligible compared with the standard gNFW mass. The data become uninformative about $L$ and any sufficiently small coupling yields essentially the same prediction for $\sigma_{\mathrm{los}}$ within the observational uncertainties, so the recovered posterior is broad and carries little information about the reference value. For $\log L_{\mathrm{ref}} \geq 0.5$, the NMC term becomes dynamically dominant and the pipeline effectively recover a non-zero coupling, but the profile is totally physically incompatible with the expected velocity profiles.

\section{Results and Discussion}
\label{sec: Discu}

The results of our statistical analysis for NGC 1052--DF2, NGC 1052--DF4, and Dragonfly 44, considering NFW and gNFW DM profiles (as representatives of the broader family of density profiles we investigated) are reported in Tables\;\ref{tab: DF2_gNFW}-\ref{tab:results_DF44_NFW}.
In these tables the list of the best-fit parameters for halo is presented: \{$c_{200}$, $M_{200}$, [$\gamma$ for gNFW]\}; stellar: \{$D$, $\Upsilon_\ast$, $v_{\rm sys}$\} and the anisotropy parameter $\beta_c$ for constant anisotropy and in case of radial anisotropy \{$\beta_0$, $\beta_\infty$, $r_a$\} alongside with the NMC parameter $L$ and Bayes factor $\log\mathcal{B}$ relative to reference models. For the two DM deficient UDGs (NGC 1052--DF2 and NGC 1052--DF4), we assume as our reference model the GR framework with only the stellar component (no DM) and constant velocity anisotropy. For the DM dominated UDG (Dragonfly 44), the reference model is still within the GR framework but includes both stars and a dark matter halo, applying the SHMR prior condition.

Before presenting the results, we note that throughout this section NFW and gNFW profiles are discussed jointly, as the two yield statistically indistinguishable outcomes across all galaxies and model configurations.

We first analyzed the systems in standard GR to build simple reference models. These GR results then serve two roles, they give us a baseline against which we can compare the NMC models, and they allow us to check if our modelling and statistical pipeline is reliable by comparing our GR results with those already published in the literature (\textit{e.g.} \cite{Bouche:2024qhy,Laudato:2022vmq,Laudato:2022drz}). 

Beyond comparing NFW and gNFW profiles, we also tested additional density profiles including Burkert, cored NFW (cNFW), Hernquist, Einasto, DARK-exp-$\gamma$, and Generalized Pseudo-Isothermal (GPI) profiles, mainly for robustness checks. In most cases they do not improve the fit quality, they do not add new elements to the final conclusions and do not alter the statistical preference with respect to the reference models for the NMC model we have considered in this work. As a result, we focus our detailed discussion on NFW and gNFW as they represent the most commonly adopted profiles in the literature, but one can see the summary of the results from other models in Table \ref{tab: All_Profiles}.

\subsection{GR baseline}

In GR, the observed line of sight velocity dispersion profiles of UDGs are explained by the combined effects of the dark matter halo mass and shape, together with the stellar orbital anisotropy. In the NMC framework, this description is extended by the introduction of a coupling scale $L$, which modifies the interaction between matter and curvature. Large values of $L$ could in principle lead to observable deviations from GR, while small values effectively reduce the theory to the GR limit.

For NGC 1052--DF2 and NGC 1052--DF4, the GR analysis recovers values fully consistent with those reported in the literature \cite{Bouche:2024qhy,Laudato:2022vmq,Laudato:2022drz}. In both systems, the inferred orbital anisotropy parameter $\beta_c$ is mostly negative, indicating a preference for tangential orbits. This behavior is robust across all explored models, while the anisotropy radius $r_a$ remains unconstrained, suggesting that the data favor a constant anisotropy description over radially varying models.

When the SHMR prior is relaxed, DF2 and DF4 admit solutions with extremely low halo masses ($\log M_{200} \sim 4$--$5\,M_\odot$) and very high concentrations ($c_{200} \sim 23$--$25$). These solutions arise from the well known degeneracy between halo mass, concentration, and anisotropy in spherical Jeans analyses \cite{Read:2017}; highly concentrated, low mass halos combined with strongly tangential orbits can reproduce the observed velocity dispersion profiles. While mathematically allowed, such configurations are astrophysically implausible, and their appearance highlights the crucial role of the SHMR prior in anchoring the halo properties to cosmologically reasonable values.

Dragonfly 44 provides a complementary case, as it is widely considered as DM dominated. In GR, Jeans modelling with the SHMR prior yields stable and physically sensible halo parameters for both NFW and gNFW profiles, with $\log M_{200} \simeq 10.9$ and concentrations $c_{200} \simeq 6$--$8$. These values remain robust against changes in the anisotropy prescription, reflecting the fact that the rising velocity-dispersion profile of DF44 tightly constrains the depth and scale of the gravitational potential. Without the SHMR prior, similar degeneracies to those seen in DF2 and DF4 reappear, but the prior effectively suppresses these non-physical regions of parameter space.

\subsection{NMC case}
\label{subsec:nmc_case}
Across all three galaxies, introducing the NMC model does not lead to significant shifts in the inferred astrophysical parameters. Distances, mass-to-light ratios, systemic velocities, and anisotropy parameters remain fully consistent with their GR counterparts within statistical uncertainties. 
The posterior distributions of the coupling length $L$ remain compatible with zero, \textit{i.e.}, with the GR limit. However, as the sensitivity analysis shows, this should not be interpreted as a tight constraint. This remains true irrespective of the presence or absence of the SHMR prior, demonstrating that the near GR outcome is not an artifact of overly flexible halo modelling.

\begin{figure*}[p]
\centering
\includegraphics[width=\textwidth,height=0.88\textheight,keepaspectratio]{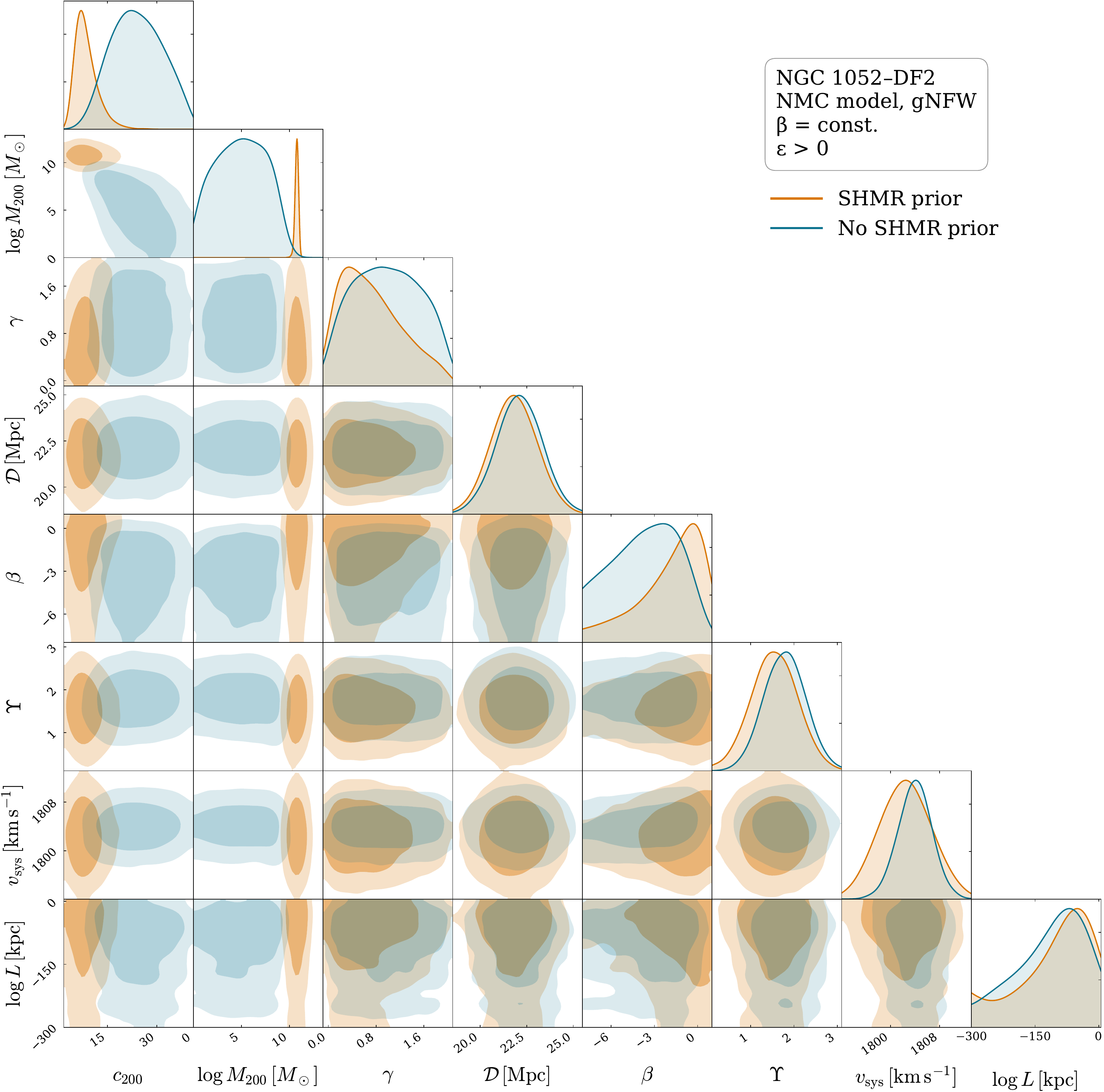}
    \caption{ Joint posterior distributions for NGC 1052--DF2 assuming a gNFW halo profile, constant velocity anisotropy, $\beta=\mathrm{const.}$, and positive coupling signature, $\varepsilon>0$. The orange contours show the case with the SHMR prior, while the teal contours show the one with NoSHMR prior. The contours indicate the $1\sigma$ and $2\sigma$ credible regions.}
    \label{fig:df2_gnfw_corner_shmr_vs_noshmr}
\end{figure*}

\begin{figure*}[p]
\centering
\includegraphics[width=\textwidth,height=0.88\textheight,keepaspectratio]{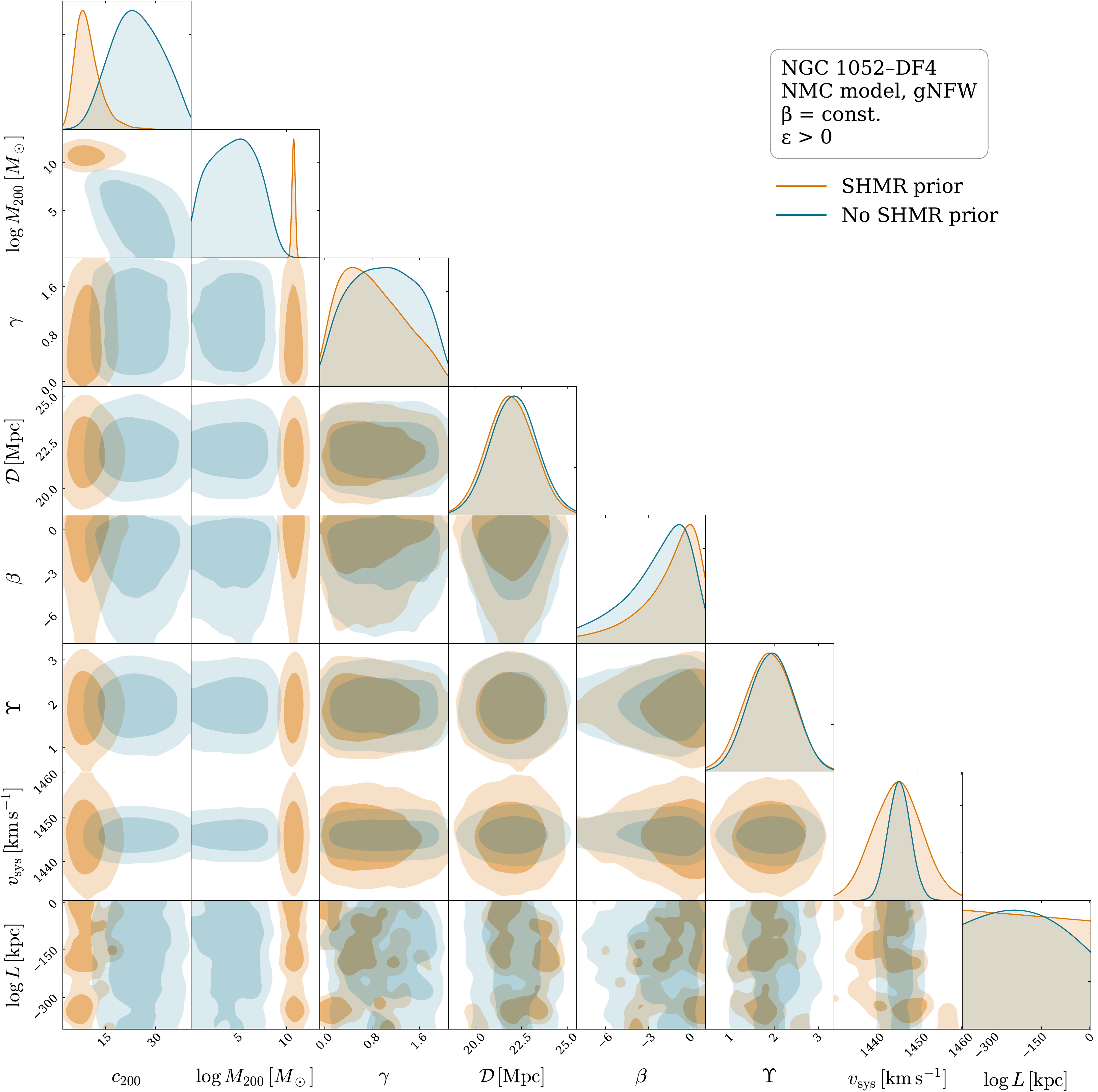}
    \caption{ Joint posterior distributions for NGC 1052--DF4 assuming a gNFW halo profile, constant velocity anisotropy, $\beta=\mathrm{const.}$, and positive coupling signature, $\varepsilon>0$. The orange contours show the run with the SHMR prior, while the teal contours show the run with NoSHMR prior. The contours indicate the $1\sigma$ and $2\sigma$ credible regions. }
    \label{fig:df4_gnfw_corner_shmr_vs_noshmr}
\end{figure*}

\begin{figure*}[p]
    \centering
    \includegraphics[width=\textwidth,height=0.88\textheight,keepaspectratio]{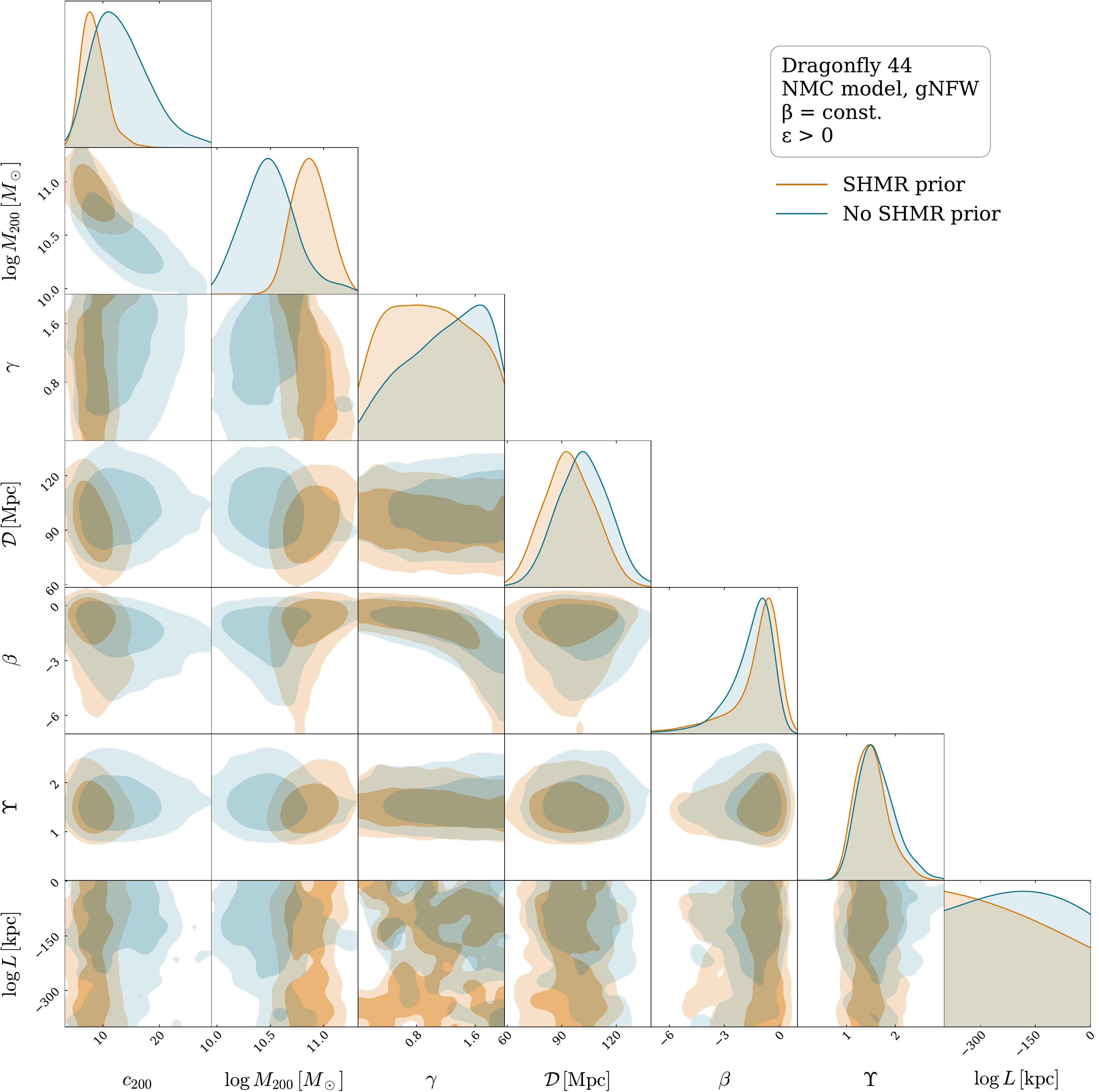}  
    \caption{Joint posterior distributions for Dragonfly 44 assuming a gNFW halo profile, constant velocity anisotropy, $\beta=\mathrm{const.}$, and positive coupling signature, $\varepsilon>0$. The orange contours show the run with the SHMR prior, while the teal contours show the run without the SHMR prior. The contours indicate the $1\sigma$ and $2\sigma$ credible regions. }
    \label{fig:df44_gnfw_corner_shmr_vs_noshmr}
\end{figure*}

\begin{figure*}[p]
    \centering
    \includegraphics[width=\textwidth,height=0.88\textheight,keepaspectratio]{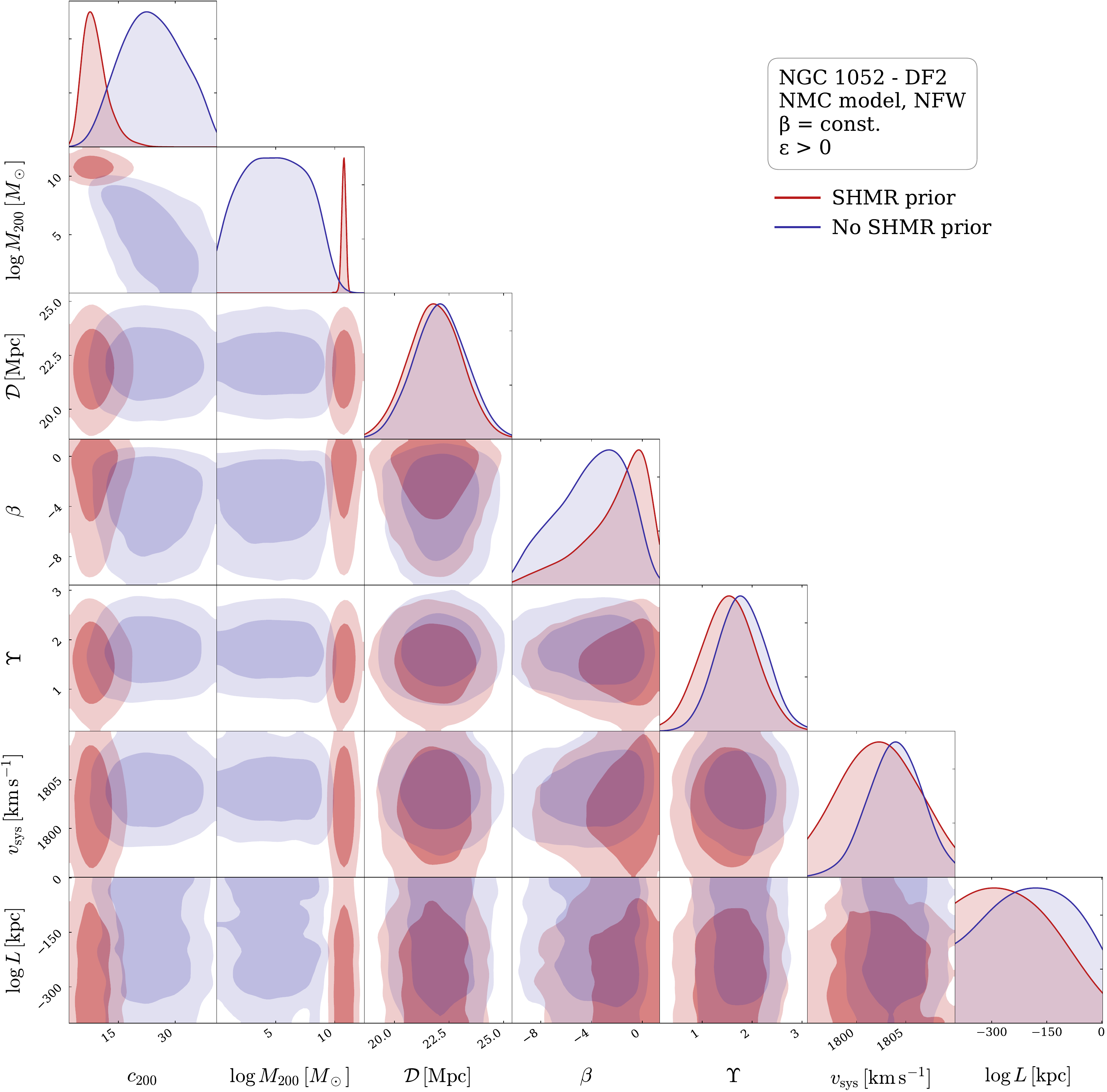}   
    \caption{Joint posterior distributions for NGC 1052--DF2 assuming an NFW halo profile, constant velocity anisotropy, $\beta=\mathrm{const.}$, and positive coupling signature, $\varepsilon>0$. The red contours show the case with the SHMR prior, while the blue contours show the run with NoSHMR prior. The contours indicate the $1\sigma$ and $2\sigma$ credible regions.}
    \label{fig:df2_nfw_corner_shmr_vs_noshmr}
\end{figure*}

\begin{figure*}[p]
    \centering
    \includegraphics[width=\textwidth,height=0.88\textheight,keepaspectratio]{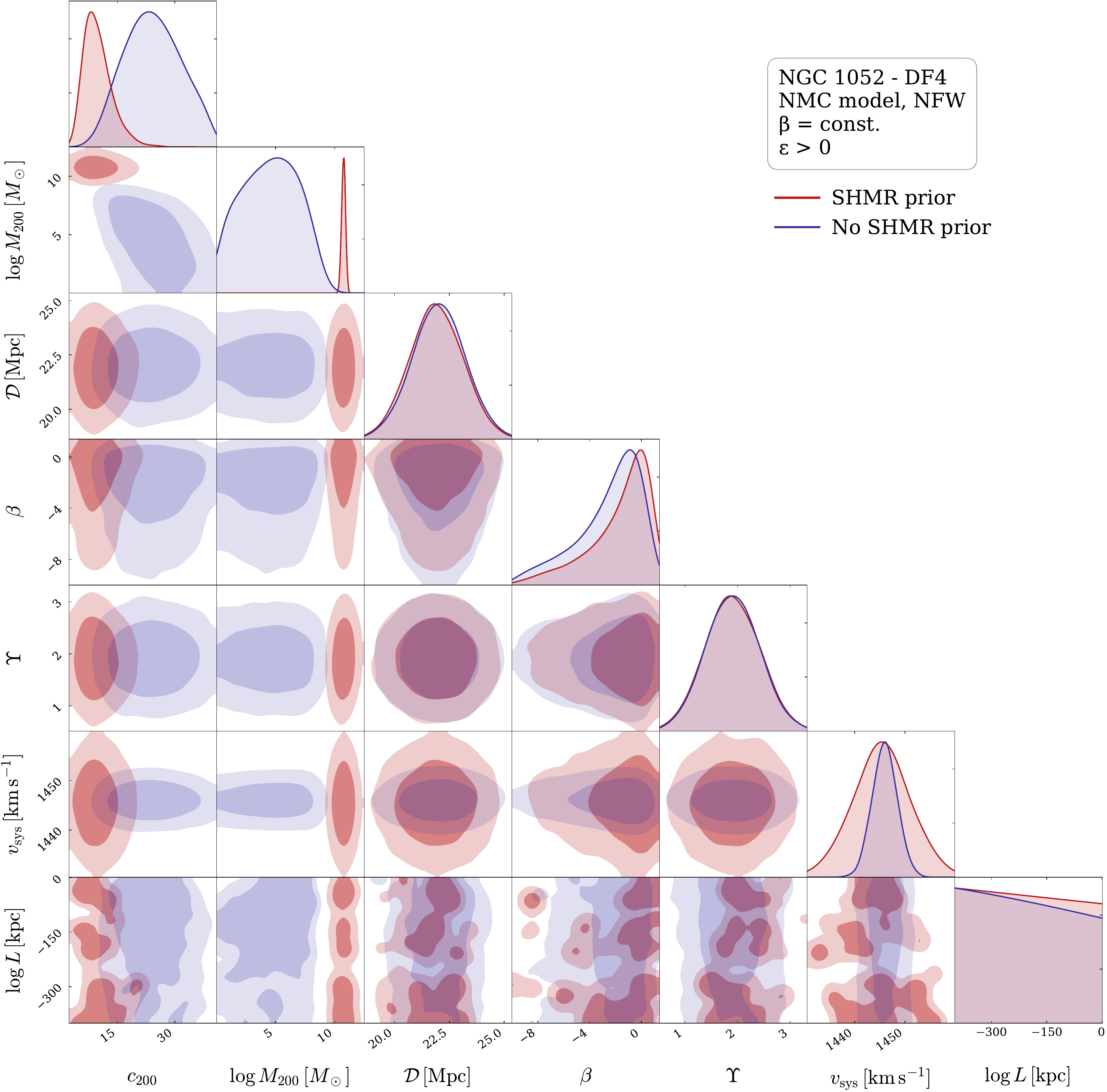}    
    \caption{ Joint posterior distributions for NGC 1052--DF4 assuming an NFW halo profile, constant velocity anisotropy, $\beta=\mathrm{const.}$, and positive coupling signature, $\varepsilon>0$. The red contours show the case with the SHMR prior, while the blue contours show the case with NoSHMR prior. The contours indicate the $1\sigma$ and $2\sigma$ credible regions.}
    \label{fig:df4_nfw_corner_shmr_vs_noshmr}
\end{figure*}

\begin{figure*}[p]
    \centering
    \includegraphics[width=\textwidth,height=0.88\textheight,keepaspectratio]{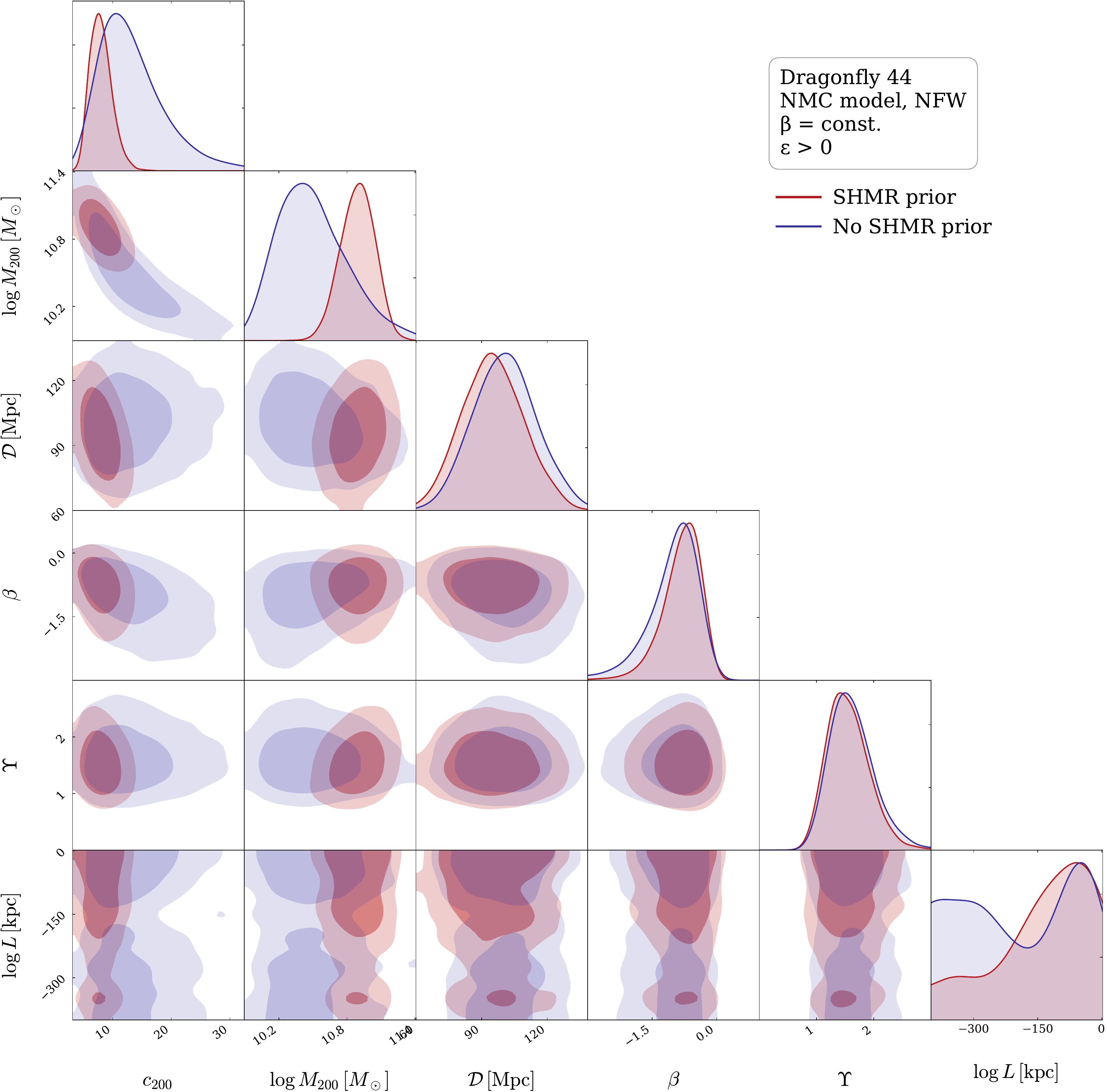}
    \caption{Joint posterior distributions for Dragonfly~44 assuming an NFW halo profile, constant velocity anisotropy, $\beta=\mathrm{const.}$, and positive coupling signature, $\varepsilon>0$. The red contours show the case with the SHMR prior, while the blue contours show the run with NoSHMR prior. The contours show the $1\sigma$ and $2\sigma$ credible regions.}
    \label{fig:df44_nfw_corner_shmr_vs_noshmr}
\end{figure*}

The corner plots in Figs. \ref{fig:df2_gnfw_corner_shmr_vs_noshmr}--\ref{fig:df44_nfw_corner_shmr_vs_noshmr} clarify the behaviour of the $L$ posteriors. As mentioned, the marginal posterior of $\log L$ is generally wide, suggesting that the data do not select a sharply preferred coupling. 
The sensitivity test of Sec. \ref{sec:Sens} demonstrates that, for $\log L_{\rm ref} \lesssim 0.5$, the NMC correction is too small to leave a detectable imprint on $\sigma_{\rm los}(R)$ within the current observational uncertainties. 

The absence of a detectable NMC signature can also be understood physically. UDGs are characterized by low masses and large sizes, leading to shallow dark matter density profiles over the radial range probed by the kinematic data. Consequently, the density varies only slowly with radius, and the curvature term $\nabla^{2}\rho_{\rm DM}$ remains small. Since the NMC correction scales as $L^{2}\nabla^{2}\rho_{\rm DM}$, its dynamical impact is strongly suppressed, even if a non-minimal interaction is present. Different dark matter profiles (cored or cuspy) therefore yield very similar constraints, as they produce nearly indistinguishable density gradients within the observed region.

The coupling scale $L$ exhibits remarkably similar behavior in all three systems, despite their very different dark matter content (DF2 and DF4 appear DM deficient, while Dragonfly 44 is DM dominated). This could suggest that, if a non-minimal coupling between dark matter and gravity exists, its effective strength on galactic scales must be small and weakly dependent on the host system. However, since the data are uninformative about $L$ in all cases, this consistency can also reflects a shared lack of sensitivity. The polarization parameter $\epsilon$ likewise has little impact, as the cases $\epsilon = \pm 1$ yield comparable constraints on $L$.

Table \ref{tab: All_Profiles} summarizes, for each halo profile, the configurations that yield the largest values of $L$, with three rows corresponding to DF2, DF4, and DF44.
For the cNFW, Einasto, and GPI profiles, no reliable constraints are reported for DF2 and DF4. The Einasto and GPI profiles are disfavored by the current UDG kinematic data, suggesting that the current observations are not sufficiently informative to constrain the extra shape parameter. For cNFW, the coupling parameter $L$ is typically unconstrained, and even within a single galaxy (and for the same polarization) the inferred $1\sigma$ values vary dramatically across runs. For this reason, we chose not to report the cNFW constraints on $L$.
Overall, the results are consistent with those we found for the NFW and gNFW profiles. The main exception is the GPI case, where DF44 shows a higher evidence and a larger value of $L$. The fact that several high $L$ selections still have negative or very small $\log\mathcal{B}$ shows that high $L$ is not preferred and our $L$ should be very small.

As mentioned, Bayesian evidence calculations confirm the conclusions drawn from the parameter posteriors. In all explored cases, the evidence ratios remain statistically inconclusive, indicating that GR and the NMC framework provide indistinguishable descriptions of the current UDG kinematic data \cite{Kass:1995loi,JeffreysScale}. For this reason, we do not report individual evidence values in the tables, as they do not add discriminatory power. Overall, the data neither favor nor disfavor the NMC model, but instead imply that any deviation from GR must be subdominant on UDG scales. Substantially more precise kinematic measurements would be required to break this degeneracy, in agreement with our previous findings at the scale of galaxy clusters \cite{Zamani:2024oep,Zamani:2024qbx}.

When looking closer to the DF44 evidence tables, some Bayesian evidence values in the NMC gNFW (Table \ref{tab: DF44_gNFW}) may require clarification. For the case including both stars and DM, with the SHMR prior and radial anisotropy, the reported $\log\mathcal{B}$ values are $0.64$ for $\epsilon<0$ and $0.65$ for $\epsilon>0$ while in GR is $-0.31$. Similar patterns occur in other cases, where posterior constraints look nearly same, yet the evidences differ noticeably. This is not abnormal because the evidence is not set only by the best fit or by the central part of the posterior. It is an integral of the likelihood over the full prior volume, so it is sensitive to how much parameter space provides a good fit. As a result, two models can look almost the same in their marginal posteriors, yet still give different $\log\mathcal{B}$. That can happen if one model has slightly broader tails, a different pattern of parameter degeneracies, or simply because the evidence estimate itself has noticeable uncertainty.

\section{Conclusions}
\label{sec: Conc}
In this work, we have explored the internal dynamics of three representative ultra-diffuse galaxies: NGC 1052--DF2, NGC 1052--DF4, and Dragonfly 44; within the framework of gravity theories featuring a non-minimal coupling between matter and curvature. Using a Bayesian Jeans analysis across multiple dark matter profiles, we aimed to test whether the kinematic data of these systems require deviations from the standard GR or not. The models were examined with the available spectroscopic data while carefully considering uncertainties in distance, stellar mass-to-light ratio, and orbital anisotropy.

\noindent Our findings show that:

\begin{itemize}[label=$\bullet$]
\item the NMC scenario provides fits statistically indistinguishable from those of GR. The inferred astrophysical parameters remain stable across the two theoretical approaches, with changes that lie within the 1$\sigma$ confidence limits. The coupling length scale $L$, which represents the strength of the non-minimal interaction, consistently yields small upper limits in all UDGs analyzed. No evidence for a significant deviation from GR is found; the posterior distributions of $L$ cluster near zero, and its effect on the velocity dispersion profiles is negligible within current observational precision.

However, the analysis presented in Sec. \ref{sec:Sens} indicates that this near-zero recovery is better understood as a consequence of the limited constraining power of the current UDG kinematics, rather than as evidence against the coupling. Specifically, the data lose sensitivity for $\log_{10}L < +0.5$, so the resulting upper limits reflect the present observational sensitivity rather than a tight exclusion of non-minimal coupling.

\item For the two galaxies with little or no dark matter (DF2 and DF4), the fits confirm that their observed motions can be explained without having a massive dark matter halo. When the stellar to halo mass relation prior is removed, the best fit halo mass drops sharply, while the concentration parameter increases, which is a behavior caused by the well known degeneracy between mass, concentration, and orbital anisotropy in Jeans analyses. In contrast, Dragonfly 44, which is dominated by dark matter, keeps a stable and physically reasonable halo even when the SHMR prior is relaxed. This highlights the very different nature of these galaxies, despite being analyzed within the same framework.

\item An important outcome based on our current work and previous studies (\cite{Zamani:2024oep,Zamani:2024qbx}) is the apparent universality of the coupling scale $L$. Despite the broad differences in mass and dark matter content among the three UDGs, our $L$ almost always stays very small. This behavior can be interpreted as if a non-minimal coupling exists in nature, its characteristic scale is likely small and nearly constant at least across galactic environments that we explored. This behavior is different from the trend reported by \cite{Gandolfi:2022puw}, who found an empirical correlation between $L$ and halo (virial) mass from stacked rotation curve analyses of disk galaxies spanning a much broader mass range. It is important to note that our analysis is based on a limited sample, including 19 galaxy clusters from our previous studies and three UDGs from the present work. Therefore, the observed trends should be interpreted with caution, and broader conclusions will require confirmation from a larger and more statistically consistent dataset.

\item The sign parameter $\epsilon$, which distinguishes between opposite coupling polarities, also remains unconstrained by current data, indicating that the existing kinematic samples lack the sensitivity to detect such delicate effects.

\end{itemize}

Overall, our study shows that ultra-diffuse galaxies, with their extreme baryon to dark matter ratios, offer a promising testing ground for gravity theories. Although, no statistically significant deviation from GR is detected here, the fact that we can see the $L$ behaving the same in both DM deficit and DM dominate environments is a key outcome. Future observations, especially those with larger samples, better velocity measurements, and more accurate distance estimates will be helpful to tighten these constraints and to determine whether NMC effects could play a measurable role in galactic dynamics. Such observations could ultimately determine whether the NMC mechanism remains a purely theoretical curiosity or emerges as a measurable feature of gravity.

We conclude our discussion with a brief look at what it would take, observationally, to place tighter constraints on the coupling scale $L$. Placing tighter constraints on $L$ would need better spectral resolution (to resolve these galaxies' small velocity dispersions), higher signal to noise (given their low surface brightness), and spatially resolved data (to map how the dispersion changes with radius, not just one average value). Deep integral field spectroscopy can complement globular cluster measurements by providing spatially resolved stellar kinematics. For DF44, stronger constraints would require higher signal to noise data and smaller uncertainties on the velocity dispersion profile. In DF2, MUSE detected a weak velocity gradient, but the resolved dispersion remains poorly constrained \cite{Emsellem:2019}, while for DF4, the Keck Cosmic Web Imager provides only an aperture integrated dispersion \cite{Shen:2023jwk}. Deeper observations could therefore improve the reconstruction of the internal dynamics and help reduce degeneracies affecting $L$. The LEWIS programme shows the potential of this approach \cite{Buttitta:2025}, although such measurements remain challenging for LSB galaxies.

\begin{acknowledgements}
The research of S.Z. and V.S. is funded by the Polish National Science Centre grant No. DEC-2021/43/O/ST9/00664. D.B. acknowledges support from grants PID2022-139841NB-I00, PID2024-158938NB-I00 funded by MICIU/AEI/10.13039/501100011033 and by “ERDF A way of making Europe”, the Project SA097P24 funded by Junta de Castilla y Le\'on.
\end{acknowledgements}

\appendix

\section{Kernel Function}
The kernel functions used in this work are adopted from \cite{Mamon:2004xk}, where $K(u,u_a) = K(\frac{r}{R},\frac{r_a}{R})$ and are given by

\begin{itemize}
    \item When $\beta$ = Const.
\end{itemize}
\be
\begin{aligned}
   K(u,u_a)= \frac{1}{2} u^{2\beta - 1} 
    &\bigg[
        \left(\frac{3}{2} - \beta \right) \sqrt{\pi} 
        \frac{\Gamma\left(\beta - \frac{1}{2}\right)}{\Gamma(\beta)} \\
       & + \beta B\left( \frac{1}{u^2}, \beta + \frac{1}{2}, \frac{1}{2} \right) \\
       & - B\left( \frac{1}{u^2}, \beta - \frac{1}{2}, \frac{1}{2} \right)
    \bigg]
\end{aligned}
\ee

$B$ is the incomplete Beta function.

\begin{itemize}
    \item when $\beta$= Radial, $u_a \neq 1$
\end{itemize}

\be
\begin{aligned}
K(u,u_a)= &\frac{1/2}{(u_a^2 - 1)}  \sqrt{1 - \frac{1}{u^2}}  + \left( 1+\frac{u_a}{u} \right) \cosh^{-1} u \\
& - \operatorname{sgn}(u_a - 1) u_a \frac{u_a^2 - 1/2}{(u_a^2 - 1)^{3/2}} \left( \frac{1 + u_a}{u} \right)\\
& \times \mathcal{C}^{-1} \left( \frac{u_a u + 1}{u + u_a} \right)
\end{aligned}
\ee

If we have
\begin{align*}
    &u_a > 1: \mathcal{C}^{-1}(x) = \cosh^{-1}x \\
   & u_a < 1: \mathcal{C}^{-1}(x) = \cos^{-1}x 
\end{align*}

\begin{itemize}
    \item when $\beta$= Radial, $u_a = 1$
\end{itemize}
\be
    \begin{aligned}
         K(u,u_a)= \left(1 + \frac{1}{u}\right) \cosh^{-1}u
- \frac{1}{6} \left( \frac{8}{u} + 7 \right) \sqrt{ \frac{u - 1}{u + 1} }
    \end{aligned}
    \ee

\section{DM halo Profiles}
\label{app: DMHalo}

Here we provide a brief overview of the halo profiles considered in our analysis.
\\

\textbf{Burkert Profile} \cite{Burkert:2000di, Mori:2000me, Salucci:2000ps}
\begin{align}
    \rho(r)&=\frac{\rho_s}{\left(1+\frac{r}{r_s}\right)\left(1+\frac{r^2}{r_s^2}\right)}\;, \\
    r_{-2}& \approx 1.521 \cdot r_s\, , \nonumber \\
    \nonumber \\ 
    M(<r) &= \pi\rho_s\,r_s^{3}\;\Big(-2\,\arctan\!\frac{r}{r_s}-4\,\ln r_s  +2\,\ln(r + r_s) \nonumber \\
    &\hspace{4.5cm} +\,\ln(r^{2} + r_s^{2})
    \Big), \nonumber \\
    \rho_s& = \frac{{\rm \Delta}}{3} \frac{\rho_{c} \left(1.521 c_{{\rm \Delta}}\right)^{3}}{-\frac{\arctan \left(1.521 c_{{\rm \Delta}}\right)}{2} 
    + \frac{\log \left[(1+(1.521 c_{{\rm \Delta}})^2)(1+1.521 c_{{\rm \Delta}})^2\right]}{4}} . \nonumber
\end{align}

\textbf{Hernquist} \cite{Hernquist:1990be,An:2012pv} 
\begin{align}
   \rho(r) & = \frac{\rho_s}{\left(\frac{r}{r_s}\right) \left(1 + \frac{r} {r_s}\right)^3}\; , \\
   r_{-2} &= \frac{r_s}{2}\, , \nonumber \\
   M(<r)&=\frac{2\pi\, \rho_s \, r^{2} \, r_{s}^{3}}{(r + r_{s})^{2}}, \nonumber\\
   \rho_{s} &= \frac{{\rm \Delta}}{3} \rho_{c}\, c_{{\rm \Delta}} \left(1+\frac{c_{{\rm \Delta}}}{2}\right)^{2}\, . \nonumber
\end{align}

\textbf{cored NFW (cNFW)} \cite{Newman:2012nv,Newman:2012nw}
\begin{align}
\rho(r) &=
\frac{\gamma \rho_s}{\left(1 + \frac{\gamma r}{r_s}\right)\left(1 + \frac{r}{r_s}\right)^{2}},\\
r_{-2} &= r_s\!\left(\frac{\tilde{\gamma}}{2\gamma}\right), \nonumber \\
M(<r) &=
\frac{4\pi\, \rho_s\,r_s^{3}\,\gamma}{(\gamma - 1)^{2}}
\Biggl[
\frac{r - r\gamma}{r + r_s}
- (\gamma - 2)\,\ln\!\left(\frac{r_s}{r + r_s}\right) \nonumber
\\[3pt]
&\qquad
+ \frac{1}{\gamma}\,\ln\!\left(1 + \frac{r\gamma}{r_s}\right)
\Biggr], \nonumber \\
\rho_s &=
\frac{\Delta}{3}\,\rho_c\,c_{\Delta}^3\,
\frac{(\gamma -1)^2 \tilde{\gamma}^3}{8 \gamma ^4 \left(
\frac{(1-\gamma)\tilde{\gamma} c_{\Delta}}{2 \gamma \left(\frac{\tilde{\gamma} c_{\Delta}}{2 \gamma }+1\right)}
+ \ln\!\frac{\left(\frac{1}{2}\tilde{\gamma} c_{\Delta}+1\right)^{\!1/\gamma}}
{\left(\frac{\tilde{\gamma} c_{\Delta}}{2 \gamma }+1\right)^{\!2-\gamma}}
\right)}. \nonumber
\end{align}

with $\tilde{\gamma} = \gamma + \sqrt{\gamma \left( 8 + \gamma \right)}$ and $\gamma>1$.\\

\textbf{Einasto} \cite{Einasto:1965czb,Retana-Montenegro:2012dbd} is defined by
\begin{align}
    \rho(r) &= \rho_s \exp\left\{-\frac{2}{\gamma} \left[\left(\frac{r}{r_s}\right)^{\gamma} - 1\right]\right\},\\
    r_{-2}&= r_s\, , \nonumber \\
    M(<r) & = 2^{\,2 - \frac{3}{\gamma}}\, \pi\, r_s^{3}\,\gamma^{-1 + \frac{3}{\gamma}}\, \exp \left[{\frac{2}{\gamma}}\right]\, \rho_s  \nonumber \\
& \qquad \cdot \left[ \Gamma\!\left(\frac{3}{\gamma}\right)-\Gamma\!\left(\frac{3}{\gamma}, \frac{2 (r/r_s)^{\gamma}}{\gamma}\right) \right], \nonumber \\
    \rho_s &= \frac{{\rm \Delta}}{3} \rho_{c} c_{{\rm \Delta}}^3 \frac{\gamma \exp\left[-\frac{2}{\gamma}\right] \left(\frac{2}{\gamma}\right)^{\frac{3}{\gamma}}}{{\rm \Gamma}\left(\frac{3}{\gamma}\right)-{\rm \Gamma}\left(\frac{3}{\gamma},\frac{2}{\gamma}c_{{\rm \Delta}}^{\gamma}\right)}\, , \nonumber
\end{align}
where ${\rm \Gamma}(z)$ and ${\rm \Gamma}(a,z)$ are respectively the Euler and incomplete gamma functions, and $0<\gamma<2$.\\

\textbf{DARK-exp-$\gamma$} \cite{Williams:2010mp,Umetsu:2015baa,Hjorth:2015bfa}
\begin{align}
        \rho(r) &= \rho_s \left(\frac{r}{r_s}\right)^{-\gamma} \left(1+\frac{r}{r_s}\right)^{\gamma-4}\; ,\\ 
        r_{-2}&= \left(1-\frac{\gamma}{2}\right)r_s\, , \nonumber \\
        M(<r)&= -\frac{4\pi\, r_s^{3}\, \rho_s}{\gamma - 3}\left(\frac{r}{r + r_s}\right)^{3 - \gamma}, \nonumber\\
        \rho_s &= \frac{{\rm \Delta}}{3}\rho_{c} c_{{\rm \Delta}}^3 \left(1-\frac{\gamma}{2}\right)^3 \frac{3-\gamma}{\frac{(1-\gamma/2) c_{{\rm \Delta}}}{1 + (1-\gamma/2) c_{{\rm \Delta}}}}\, \nonumber,
\end{align}

with $0<\gamma<2$.
\vspace{0.3cm}

\textbf{Generalized Pseudo Isothermal (GPI)} \cite{Begeman:1991iy}
\begin{align}
\rho(r) & = \frac{\rho_s}{\left[1 + \left(\frac{r}{r_s}\right)^2\right]^\gamma}\, ,
\\
r_{-2} & = \frac{r_s}{\sqrt{\gamma-1}}\, , \nonumber \\
M(<r)&=\frac{4\pi\,\rho_s\,r_s^{2\gamma}\,(r^2 + r_s^2)^{1 - \gamma}} {r\,[3 + 4(\gamma - 2)\gamma]}\!
        \Biggl[
          r^{2}(1 - 2\gamma) - r_s^{2} \notag \\
&\qquad\qquad
          + r_s^{2}\,{}_2F_1\!\left[1, \tfrac{1}{2} - \gamma;\, \tfrac{1}{2};\, -\frac{r^{2}}{r_s^{2}}\right]
        \Biggr], \nonumber \\
\rho_s &= \frac{{\rm \Delta}}{3} \rho_{c}
\frac{\frac{c^{4}_{{\rm \Delta}}}{\left(\gamma-1\right)^2}}
{\left[ 1 + \frac{c^{2}_{{\rm \Delta}}}{\gamma-1} \right]^{\gamma-1}} \nonumber \\
&\cdot \frac{3+4(\gamma-2)\gamma}
{-1+\frac{(1-2\gamma)}{\gamma-1}c^{2}_{{\rm \Delta}}+{_2}F_{1} \left[1,\frac{1}{2}-\gamma;\frac{1}{2};-\frac{c^{2}_{{\rm \Delta}}}{\gamma-1}\right]}\, , \nonumber
\end{align}
with $\gamma>1$.

{\renewcommand{\tabcolsep}{2.5mm}
{\renewcommand{\arraystretch}{2.5}
\begin{table*}
\begin{minipage}{\textwidth}
\centering
\caption{Results from the statistical analysis of NGC1052--DF2. considering gNFW profile}
\label{tab: DF2_gNFW}
\resizebox*{\textwidth}{!}{
\begin{tabular}{c|cccc|cccc|ccc|c}
\tightdoublehline
 & \multicolumn{12}{c}{GR}   \\
\hline
& $c_{200}$ & $\log M_{200}$ & $\gamma$ & $\log L$ & $\beta_c$ & $\beta_0$ & $\beta_{\infty}$ & $r_a$ & $D$ & $\Upsilon_{\ast}$ & $v_{sys}$ & $\log\mathcal{B}^{i}_{j}$ \\
 &  & M$_{\odot}$ &  & kpc  &  &  &  & kpc & Mpc &  &  km s$^{-1}$ &  \\ 
\hline
\multirow{2}{*}{Stars +  DM (SHMR)} & $8.14^{+3.75}_{-2.46}$ & $10.78^{+0.16}_{-0.18}$ & $0.66^{+0.61}_{-0.45}$ & $-$ & $-1.51^{+1.68}_{-3.55}$ & $-$ & $-$ & $-$ & $21.87^{+1.23}_{-1.20}$ & $1.56^{+0.51}_{-0.52}$ & $1802.31^{+4.20}_{-4.33}$  & $-0.24^{+0.02}_{-0.02}$  \\
& $7.69^{+3.53}_{-2.34}$ & $10.77^{+0.17}_{-0.18}$ & $0.65^{+0.64}_{-0.43}$ & $-$ & $-$ & $-1.00^{+0.98}_{-1.80}$ & $-0.50^{+0.84}_{-1.72}$ & $U$ & $21.82^{+1.22}_{-1.20}$ & $1.52^{+0.52}_{-0.52}$ & $1801.88^{+4.24}_{-4.41}$ & $-0.56^{+0.02}_{-0.02}$\\
\hline
\multirow{2}{*}{Stars + DM (NoSHMR)} & $23.21^{+8.63}_{-7.44}$ & $4.95^{+2.81}_{-2.93}$ & $1.01^{+0.61}_{-0.62}$ & $-$ & $-3.67^{+2.37}_{-3.21}$ & $-$ & $-$ & $-$ & $22.11^{+1.18}_{-1.22}$ & $1.79^{+0.49}_{-0.46}$ & $1803.95^{+2.63}_{-2.64}$ & $-0.16^{+0.01}_{-0.01}$ \\
& $24.46^{+8.03}_{-7.54}$ & $4.45^{+2.62}_{-2.64}$ & $U$ & $-$ & $-$ & $-3.45^{+1.94}_{-2.88}$ & $-1.17^{+1.22}_{-2.35}$ & $U$ & $22.07^{+1.20}_{-1.18}$ & $1.64^{+0.49}_{-0.48}$ & $1804.25^{+3.01}_{-3.03}$ & $-0.50^{+0.01}_{-0.01}$ \\
\hline
\multirow{1}{*}{Stars} & $-$ & $-$ & $-$ & $-$ & $-3.77^{+2.38}_{-3.22}$ & $-$ & $-$ & $-$ & $22.12^{+1.18}_{-1.18}$ & $1.81^{+0.46}_{-0.44}$ & $1804.03^{+2.58}_{-2.64}$ & $0$ \\
\tightdoublehline
 & \multicolumn{11}{c}{Non Minimal Coupling with $\epsilon = -1$}   \\
\hline
  & $c_{200}$ & $\log M_{200}$ & $\gamma$ & $\log L$ & $\beta_c$ & $\beta_0$ & $\beta_{\infty}$ & $r_a$ & $D$ & $\Upsilon_{\ast}$ & $v_{sys}$ & $\log\mathcal{B}^{i}_{j}$ \\
  & & M$_{\odot}$ &  & kpc  &  &  &  & kpc & Mpc &  &  km s$^{-1}$  \\ 
\hline
\multirow{2}{*}{Stars + DM (SHMR)} & $8.01^{+3.70}_{-2.36}$ & $10.78^{+0.17}_{-0.18}$ & $>0.19$ & $<1.18$ & $<0.30$ & $-$ & $-$ & $-$ & $21.82^{+1.23}_{-1.17}$ & $1.55^{+0.51}_{-0.52}$ & $1802.28^{+4.10}_{-4.28}$ & $-0.16^{+0.02}_{-0.02}$ \\
& $7.89^{+3.64}_{-2.49}$ & $10.77^{+0.16}_{-0.18}$ & $0.64^{+0.65}_{-0.41}$ & $<0.10$ & $-$ & $<0.00$ & $<0.35$ & $U$ & $21.86^{+1.25}_{-1.21}$ & $1.54^{+0.52}_{-0.53}$ & $1801.93^{+4.29}_{-4.30}$ & $-0.60^{+0.01}_{-0.02}$ \\
\hline
\multirow{2}{*}{Stars + DM (NoSHMR)} & $23.37^{+8.48}_{-7.36}$ & $4.87^{+2.85}_{-2.89}$ & $U$ & $<0.95$ & $-3.83^{+2.45}_{-3.18}$ & $-$ & $-$ & $-$ & $22.14^{+1.19}_{-1.20}$ & $1.77^{+0.47}_{-0.46}$ & $1804.01^{+2.69}_{-2.70}$ & $-0.14^{+0.01}_{-0.01}$ \\
& $24.83^{+8.27}_{-7.84}$ & $4.35^{+2.50}_{-2.54}$ & $U$ & $<1.74$ & $-$ & $-3.60^{+2.02}_{-2.84}$ & $<0.13$ & $U$ & $22.08^{+1.21}_{-1.21}$ & $1.65^{+0.48}_{-0.46}$ & $1804.23^{+3.02}_{-3.08}$ & $-0.52^{+0.01}_{-0.01}$  \\
\tightdoublehline
 & \multicolumn{11}{c}{Non Minimal Coupling with $\epsilon = +1$}   \\
\hline
  & $c_{200}$ & $\log M_{200}$ & $\gamma$ & $\log L$ & $\beta_c$ & $\beta_0$ & $\beta_{\infty}$ & $r_a$ & $D$ & $\Upsilon_{\ast}$ & $v_{sys}$ & $\log\mathcal{B}^{i}_{j}$   \\
  & & M$_{\odot}$ &  & kpc  &  &  &  & kpc & Mpc &  &  km s$^{-1}$ \\ 
\hline
\multirow{2}{*}{Stars+DM (SHMR)} & $7.95^{+3.81}_{-2.37}$ & $10.78^{+0.17}_{-0.18}$ & $0.65^{+0.64}_{-0.44}$ & $<-0.47$ & $<0.13$ & $-$ & $-$ & $-$ & $21.81^{+1.22}_{-1.19}$ & $1.55^{+0.52}_{-0.52}$ & $1802.43^{+4.17}_{-4.16}$ & $-0.20^{+0.01}_{-0.01}$ \\
& $7.69^{+3.37}_{-2.34}$ & $10.77^{+0.17}_{-0.18}$ & $>0.18$ & $<-7.37$ & $-$ & $<0.11$ & $<0.41$ & $U$ & $21.87^{+1.24}_{-1.21}$ & $1.54^{+0.53}_{-0.53}$ & $1802.03^{+4.10}_{-4.26}$ & $-0.59^{+0.02}_{-0.01}$  \\
\hline
\multirow{2}{*}{Stars+DM (NoSHMR)} & $23.76^{+8.54}_{-7.71}$ & $4.94^{+2.82}_{-2.97}$ & $U$ & $<-2.77$ & $-3.75^{+2.45}_{-3.26}$ & $-$ & $-$ & $-$ & $22.12^{+1.17}_{-1.18}$ & $1.79^{+0.47}_{-0.47}$ & $1804.00^{+2.61}_{-2.71}$ & $-0.13^{+0.01}_{-0.01}$\\
& $24.41^{+8.15}_{-7.59}$ & $4.37^{+2.62}_{-2.61}$ & $U$ & $<0.23$ & $-$ & $-3.42^{+1.95}_{-2.95}$ & $<0.15$ & $U$ & $22.11^{+1.20}_{-1.22}$ & $1.63^{+0.49}_{-0.48}$ & $1804.26^{+2.91}_{-3.04}$ & $-0.49^{+0.01}_{-0.01}$\\
\tightdoublehline
\end{tabular}}
\end{minipage}
\end{table*}}}

{\renewcommand{\tabcolsep}{2.5mm}
{\renewcommand{\arraystretch}{2.5}
\begin{table*}
\begin{minipage}{\textwidth}
\centering
\caption{Results from the statistical analysis of NGC1052-DF4. considering gNFW profile}
\label{tab: DF4_gNFW}
\resizebox*{\textwidth}{!}{
\begin{tabular}{c|cccc|cccc|ccc|c}
\tightdoublehline
 & \multicolumn{12}{c}{GR}   \\
\hline
  & $c_{200}$ & $\log M_{200}$ & $\gamma$ & $\log L$ & $\beta_c$ & $\beta_0$ & $\beta_{\infty}$ & $r_a$ & $D$ & $\Upsilon_{\ast}$ & $v_{sys}$ & $\log\mathcal{B}^{i}_{j}$ \\
 &  & M$_{\odot}$ &  & kpc  &  &  &  & kpc & Mpc &  &  km s$^{-1}$ & \\ 
\hline
\multirow{2}{*}{Stars +  DM (SHMR)} & $9.06^{+4.12}_{-2.72}$ & $10.75^{+0.16}_{-0.17}$ & $0.73^{+0.61}_{-0.47}$ & $-$ & $<0.46$ & $-$ & $-$ & $-$ & $21.93^{+1.20}_{-1.21}$ & $1.90^{+0.51}_{-0.50}$ & $1445.67^{+5.30}_{-5.60}$ & $-0.12^{+0.01}_{-0.02}$\\
& $8.84^{+4.28}_{-2.67}$ & $10.75^{+0.16}_{-0.17}$ & $0.71^{+0.62}_{-0.47}$ & $-$ & $-$ & $<0.03$ & $<0.42$ & $U$ & $21.92^{+1.18}_{-1.20}$ & $1.89^{+0.51}_{-0.52}$ & $1445.04^{+5.98}_{-6.15}$ & $-0.62^{+0.02}_{-0.02}$  \\
\hline
\multirow{2}{*}{Stars + DM (NoSHMR)} & $24.39^{+8.33}_{-7.68}$ & $4.47^{+2.59}_{-2.69}$ & $1.00^{+0.61}_{-0.62}$ & $-$ & $<0.96$ & $-$ & $-$ & $-$ & $22.05^{+1.20}_{-1.20}$ & $1.93^{+0.49}_{-0.49}$ & $1445.95^{+2.44}_{-2.44}$ & $-0.12^{+0.02}_{-0.02}$\\
& $24.59^{+7.86}_{-7.53}$ & $4.22^{+2.44}_{-2.57}$ & $1.00^{+0.61}_{-0.61}$ & $-$ & $-$ & $<-0.65$ & $<-0.27$ & $U$ & $21.99^{+1.18}_{-1.20}$ & $1.83^{+0.50}_{-0.53}$ & $1445.72^{+3.03}_{-3.10}$ & $-0.55^{+0.01}_{-0.01}$\\
\hline
\multirow{1}{*}{Stars} & $-$ & $-$ & $-$ & $-$ & $<-0.99$ & $-$ & $-$ & $-$ & $22.04^{+1.23}_{-1.19}$ & $1.91^{+0.49}_{-0.48}$ & $1445.94^{+2.37}_{-2.38}$ & $0$ \\
\tightdoublehline
 & \multicolumn{12}{c}{Non Minimal Coupling with $\epsilon = -1$}   \\
\hline
  & $c_{200}$ & $\log M_{200}$ & $\gamma$ & $\log L$ & $\beta_c$ & $\beta_0$ & $\beta_{\infty}$ & $r_a$ & $D$ & $\Upsilon_{\ast}$ & $v_{sys}$ & $\log\mathcal{B}^{i}_{j}$  \\
  & & M$_{\odot}$ &  & kpc  &  &  &  & kpc & Mpc &  &  km s$^{-1}$  \\ 
\hline
\multirow{2}{*}{Stars + DM (SHMR)} & $9.16^{+4.09}_{-2.78}$ & $10.75^{+0.17}_{-0.17}$ & $0.74^{+0.62}_{-0.48}$ & $<-5.14$ & $<0.48$ & $-$ & $-$ & $-$ & $21.89^{+1.19}_{-1.22}$ & $1.89^{+0.51}_{-0.51}$ & $1445.36^{+5.31}_{-5.30}$ & $-0.11^{+0.02}_{-0.01}$ \\ 
& $8.77^{+3.94}_{-2.74}$ & $10.75^{+0.16}_{-0.17}$ & $>0.20$ & $<0.14$ & $-$ & $<0.04$ & $<0.42$ & $U$ & $21.89^{+1.19}_{-1.19}$ & $1.89^{+0.52}_{-0.53}$ & $1445.4^{+5.96}_{-6.24}$ & $-0.54^{+0.02}_{-0.01}$  \\
\hline
\multirow{2}{*}{Stars + DM (NoSHMR)} & $24.35^{+8.04}_{-7.48}$ & $4.51^{+2.52}_{-2.66}$ & $U$ & $<-0.01$ & $<-0.12$ & $-$ & $-$ & $-$ & $22.04^{+1.20}_{-1.19}$ & $1.91^{+0.50}_{-0.49}$ & $1445.93^{+2.51}_{-2.44}$ & $-0.13^{+0.02}_{-0.01}$ \\ 
& $24.60^{+8.07}_{-7.40}$ & $4.16^{+2.51}_{-2.56}$ & $U$ & $<-6.17$ & $-$ & $<-0.59$ & $<0.31$ & $U$ & $22.04^{+1.20}_{-1.19}$ & $1.82^{+0.51}_{-0.51}$ & $1445.59^{+2.98}_{-3.04}$  & $-0.51^{+0.02}_{-0.01}$  \\
\tightdoublehline
 & \multicolumn{12}{c}{Non Minimal Coupling with $\epsilon = +1$}   \\
\hline
  & $c_{200}$ & $\log M_{200}$ & $\gamma$ & $\log L$ & $\beta_c$ & $\beta_0$ & $\beta_{\infty}$ & $r_a$ & $D$ & $\Upsilon_{\ast}$ & $v_{sys}$ & $\log\mathcal{B}^{i}_{j}$ \\
  & & M$_{\odot}$ &  & kpc  &  &  &  & kpc & Mpc &  &  km s$^{-1}$ &   \\ 
\hline
\multirow{2}{*}{Stars+DM (SHMR)} & $9.11^{+4.12}_{-2.71}$ & $10.75^{+0.16}_{-0.17}$ & $0.74^{+0.65}_{-0.48}$ & $<-9.24$ & $<0.47$ & $-$ & $-$ & $-$ & $21.91^{+1.25}_{-1.22}$ & $1.89^{+0.53}_{-0.53}$ & $1445.62^{+5.30}_{-5.57}$ & $-0.12^{+0.02}_{-0.02}$ \\ 
& $8.92^{+4.01}_{-2.74}$ & $10.75^{+0.16}_{-0.16}$ & $>0.21$ & $<-109.05$ & $-$ & $<0.01$ & $<0.41$ & $U$ & $21.87^{+1.19}_{-1.14}$ & $1.90^{+0.51}_{-0.52}$ & $1445.34^{+5.92}_{-6.14}$ & $-0.59^{+0.02}_{-0.01}$ \\
\hline
\multirow{2}{*}{Stars+DM (NoSHMR)} & $24.33^{+8.10}_{-7.34}$ & $4.43^{+2.55}_{-2.74}$ & $1.94^{+0.50}_{-0.50}$ & $<-6.93$ & $<-0.13$ & $-$ & $-$ & $-$ & $22.10^{+1.21}_{-1.21}$ & $1.94^{+0.50}_{-0.50}$ & $1445.93^{+2.46}_{-2.43}$ & $-0.14^{+0.02}_{-0.01}$ \\ 
& $24.50^{+8.11}_{-7.37}$ & $4.14^{+2.53}_{-2.50}$ & $1.01^{+0.60}_{-0.62}$ & $<-22.83$ & $-$ & $<-0.63$ & $<0.28$ & $U$ & $21.98^{+1.19}_{-1.15}$ & $1.81^{+0.51}_{-0.50}$ & $1445.62^{+2.90}_{-2.98}$ & $-0.52^{+0.02}_{-0.01}$ \\
\tightdoublehline
\end{tabular}}
\end{minipage}
\end{table*}}}

{\renewcommand{\tabcolsep}{2.5mm}
{\renewcommand{\arraystretch}{2.5}
\begin{table*}
\begin{minipage}{\textwidth}
\centering
\caption{Results from the statistical analysis of Dragonfly 44, considering gNFW profile}
\label{tab: DF44_gNFW}
\resizebox*{\textwidth}{!}{
\begin{tabular}{c|cccc|cccc|cc|c}
\tightdoublehline
 & \multicolumn{11}{c}{GR}   \\
\hline
  & $c_{200}$ & $\log M_{200}$ & $\gamma$ & $\log L$ & $\beta_c$ & $\beta_0$ & $\beta_{\infty}$ & $r_a$ & $D$ & $\Upsilon_{\ast}$ & $\log\mathcal{B}^{i}_{j}$ \\
 &  & M$_{\odot}$ &  & kpc  &  &  &  & kpc & Mpc & &  \\ 
\hline
\multirow{2}{*}{Stars +  DM (SHMR)} & $8.12^{+2.30}_{-1.78}$ & $10.87^{+0.17}_{-0.17}$ & $U$ & $-$ & $<-0.22$ & $-$ & $-$ & $-$ & $93.62^{+13.56}_{-12.90}$ & $1.45^{+0.39}_{-0.28}$ & $0$ \\
& $6.50^{+1.86}_{-1.44}$ & $10.88^{+0.15}_{-0.17}$ & $>0.20$ & $-$ & $-$ & $<-0.48$ & $<-0.07$ & $U$ & $89.83^{+13.54}_{-13.45}$ & $1.48^{+0.38}_{-0.30}$ & $-0.31^{+0.02}_{-0.02}$ \\
\hline
\multirow{2}{*}{Stars + DM (NoSHMR)} & $12.55^{+6.87}_{-5.20}$ & $10.51^{+0.35}_{-0.26}$ & $U$ & $-$ & $<-0.34$ & $-$ & $-$ & $-$ & $101.29^{+14.15}_{-13.52}$ & $1.61^{+0.39}_{-0.36}$ & $0.21^{+0.02}_{-0.02}$  \\
& $11.17^{+5.83}_{-3.70}$ & $10.05^{+0.37}_{-0.34}$ & $<1.01$ & $-$ & $-$ & $<-1.96$ & $<-1.07$ & $U$ & $100.00^{+14.20}_{-14.34}$ & $1.61^{+0.43}_{-0.34}$ & $-0.07^{+0.02}_{-0.02}$   \\
\tightdoublehline
 & \multicolumn{11}{c}{Non Minimal Coupling with $\epsilon = -1$}   \\
\hline
  & $c_{200}$ & $\log M_{200}$ & $\gamma$ & $\log L$ & $\beta_c$ & $\beta_0$ & $\beta_{\infty}$ & $r_a$ & $D$ & $\Upsilon_{\ast}$ & $\log\mathcal{B}^{i}_{j}$ \\
  & & M$_{\odot}$ &  & kpc  &  &  &  & kpc & Mpc & & \\ 
\hline
\multirow{2}{*}{Stars + DM (SHMR)} & $7.97^{+2.36}_{-1.60}$ & $10.87^{+0.17}_{-0.16}$ & $1.01^{+0.64}_{-0.60}$ & $<-0.27$ & $<-0.27$ & $-$ & $-$ & $-$ & $93.58^{+13.75}_{-13.48}$ & $1.49^{+0.39}_{-0.32}$ & $-0.00^{+0.02}_{-0.01}$  \\ 
& $6.45^{+1.90}_{-1.52}$ & $10.87^{+0.16}_{-0.16}$ & $0.57^{+0.47}_{-0.36}$ & $<-10.65$ & $-$ & $<-0.47$ & $<-0.07$ & $U$ & $89.56^{+14.46}_{-14.01}$ & $1.49^{+0.38}_{-0.31}$ & $0.64^{+0.02}_{-0.02}$ \\
\hline
\multirow{2}{*}{Stars + DM (NoSHMR)} & $12.62^{+5.90}_{-4.02}$ & $10.49^{+0.25}_{-0.23}$ & $U$ & $<-1.23$ & $<-0.47$ & $-$ & $-$ & $-$ & $101.60^{+15.43}_{-13.97}$ & $1.60^{+0.41}_{-0.33}$ & $0.21^{+0.02}_{-0.01}$ \\ 
& $12.84^{+12.21}_{-4.49}$ & $9.90^{+0.41}_{-7.09}$ & $0.79^{+0.58}_{-0.48}$ & $<-0.45$ & $-$ & $-3.62^{+2.03}_{-4.28}$ & $<-0.05$ & $U$ & $103.05^{+15.14}_{-15.10}$ & $1.75^{+0.61}_{-0.42}$ & $-0.20^{+0.02}_{-0.02}$\\
\tightdoublehline
 & \multicolumn{11}{c}{Non Minimal Coupling with $\epsilon = +1$}   \\
\hline
  & $c_{200}$ & $\log M_{200}$ & $\gamma$ & $\log L$ & $\beta_c$ & $\beta_0$ & $\beta_{\infty}$ & $r_a$ & $D$ & $\Upsilon_{\ast}$ & $\log\mathcal{B}^{i}_{j}$ \\
  & & M$_{\odot}$ &  & kpc  &  &  &  & kpc & Mpc & & \\ 
\hline
\multirow{2}{*}{Stars+DM (SHMR)} & $8.19^{+2.31}_{-1.84}$ & $10.87^{+0.18}_{-0.16}$ & $U$ & $<-8.22$ & $<-0.26$ & $-$ & $-$ & $-$ & $93.61^{+14.82}_{-13.72}$ & $1.49^{+0.34}_{-0.31}$ & $0.04^{+0.01}_{-0.02}$  \\ 
& $6.55^{+1.76}_{-1.52}$ & $10.87^{+0.16}_{-0.17}$ & $0.60^{+0.48}_{-0.38}$ & $<-2.10$ & $-$ & $<-0.51$ & $<-0.11$ & $U$ & $90.30^{+14.42}_{-13.64}$ & $1.48^{+0.38}_{-0.31}$ & $0.65^{+0.02}_{-0.02}$ \\
\hline
\multirow{2}{*}{Stars+DM (NoSHMR)} & $12.68^{+5.63}_{-4.07}$ & $10.48^{+0.24}_{-0.23}$ & $<1.80$ & $<-0.13$ & $<-0.51$ & $-$ & $-$ & $-$ & $101.71^{+14.23}_{-14.00}$ & $1.57^{+0.42}_{-0.31}$ & $0.28^{+0.01}_{-0.02}$ \\ 
& $15.95^{+12.20}_{-7.31}$ & $9.41^{+0.84}_{-6.44}$ & $U$ & $<-0.96$ & $-$ & $U$ & $<0.24$ & $U$ & $104.93^{+14.77}_{-15.08}$ & $1.91^{+0.48}_{-0.51}$ & $0.67^{+0.03}_{-0.03}$ \\
\tightdoublehline
\end{tabular}}
\end{minipage}
\end{table*}}}

{\renewcommand{\tabcolsep}{2.5mm}
{\renewcommand{\arraystretch}{2.5}
\begin{table*}
\begin{minipage}{\textwidth}
\centering
\caption{Results from the statistical analysis of NGC1052-DF2. considering NFW profile}\label{tab:results_DF2_NFW}
\resizebox*{\textwidth}{!}{
\begin{tabular}{c|ccc|cccc|ccc|c}
\tightdoublehline
& \multicolumn{10}{c}{GR}   \\
\hline
  & $c_{200}$ & $\log M_{200}$ & $\log L$ & $\beta_c$ & $\beta_0$ & $\beta_{\infty}$ & $r_a$ & $D$ & $\Upsilon_{\ast}$ & $v_{sys}$ & $\log\mathcal{B}^{i}_{j}$ \\
  &  & M$_{\odot}$ & kpc &  &  &  & kpc & Mpc & & km s$^{-1}$ &   \\ 
\hline
\multirow{2}{*}{Stars + DM (SHMR)} 
 & $8.34^{+3.79}_{-2.45}$ & $10.77^{+0.16}_{-0.18}$ & $-$ & $<-0.40$ & $-$ & $-$ & $-$ & $21.87^{+1.21}_{-1.24}$ & $1.55^{+0.52}_{-0.52}$ & $1802.31^{+4.43}_{-4.30}$ & $-0.40^{+0.01}_{-0.01}$ \\
 & $8.15^{+3.66}_{-2.44}$ & $10.76^{+0.16}_{-0.19}$ & $-$ & $-$ & $<-0.51$ & $<-0.07$ & $U$ & $21.85^{+1.20}_{-1.19}$ & $1.52^{+0.52}_{-0.51}$ & $1802.23^{+4.33}_{-4.41}$ & $0.23^{+0.02}_{-0.02}$ \\
\hline
\multirow{2}{*}{Stars + DM (NoSHMR)} 
 & $23.35^{+8.32}_{-7.83}$ & $5.03^{+2.78}_{-2.97}$ & $-$ & $-3.81^{+2.45}_{-3.19}$ & $-$ & $-$ & $-$ & $22.11^{+1.22}_{-1.19}$ & $1.80^{+0.43}_{-0.46}$ & $1804.03^{+2.64}_{-2.72}$ & $-0.15^{+0.01}_{-0.01}$ \\
 & $24.32^{+8.18}_{-7.49}$ & $4.32^{+2.60}_{-2.62}$ & $-$ & $-$ & $-3.54^{+1.99}_{-2.81}$ & $<-0.50$ & $U$ & $22.09^{+1.21}_{-1.19}$ & $1.66^{+0.49}_{-0.49}$ & $1804.26^{+2.96}_{-3.22}$ & $-0.51^{+0.01}_{-0.02}$\\
\hline
\multirow{1}{*}{Stars} 
 & $-$ & $-$ & $-$ & $-3.77^{+2.38}_{-3.22}$ & $-$ & $-$ & $-$ & $22.12^{+1.18}_{-1.18}$ & $1.81^{+0.46}_{-0.45}$ & $1804.03^{+2.58}_{-2.64}$ & $0$  \\
\tightdoublehline
 & \multicolumn{10}{c}{Non Minimal Coupling with $\epsilon = -1$}   \\
\hline
  & $c_{200}$ & $\log M_{200}$ & $\log L$ & $\beta_c$ & $\beta_0$ & $\beta_{\infty}$ & $r_a$ & $D$ & $\Upsilon_{\ast}$ & $v_{sys}$ & $\log\mathcal{B}^{i}_{j}$ \\
  & & M$_{\odot}$  & kpc  &  &  &  & kpc & Mpc &  &  km s$^{-1}$ & \\ 
\hline
\multirow{2}{*}{Stars + DM (SHMR)} & $8.25^{+3.59}_{-2.37}$ & $10.77^{+0.16}_{-0.18}$ & $<0.05$ & $<0.29$ & $-$ & $-$ & $-$ & $21.87^{+1.18}_{-1.19}$ & $1.52^{+0.51}_{-0.52}$ & $1802.23^{+4.30}_{-4.29}$ & $-0.36^{+0.01}_{-0.02}$ \\ 
 & $7.95^{+3.50}_{-2.30}$ & $10.76^{+0.16}_{-0.18}$ & $<-5.27$ & $-$ & $<0.06$ & $<0.39$ & $U$ & $21.84^{+1.20}_{-1.20}$ & $1.53^{+0.52}_{-0.52}$ & $1802.03^{+4.45}_{-4.52}$ & $0.20^{+0.02}_{-0.02}$  \\
\hline
\multirow{2}{*}{Stars + DM (NoSHMR)} & $23.13^{+8.72}_{-7.79}$ & $4.97^{+2.91}_{-2.99}$ & $<2.10$ & $-3.80^{+2.47}_{-3.30}$ & $-$ & $-$ & $-$ & $22.14^{+1.18}_{-1.22}$ & $1.78^{+0.49}_{-0.46}$ & $1803.99^{+2.74}_{-2.78}$ & $-0.16^{+0.01}_{-0.02}$\\ 
 & $24.37^{+8.53}_{-7.41}$ & $4.30^{+2.62}_{-2.58}$ & $<-25.27$ & $-$ & $-3.54^{+1.99}_{-2.82}$ & $<0.15$ & $U$ & $22.14^{+1.18}_{-1.22}$ & $1.63^{+0.48}_{-0.47}$ & $1804.23^{+2.93}_{-3.07}$ & $-0.48^{+0.01}_{-0.01}$ \\
\tightdoublehline
 & \multicolumn{10}{c}{Non Minimal Coupling with $\epsilon = +1$}   \\
\hline
  & $c_{200}$ & $\log M_{200}$ & $\log L$ & $\beta_c$ & $\beta_0$ & $\beta_{\infty}$ & $r_a$ & $D$ & $\Upsilon_{\ast}$ & $v_{sys}$ & $\log\mathcal{B}^{i}_{j}$\\
  & & M$_{\odot}$  & kpc  &  &  &  & kpc & Mpc &  &  km s$^{-1}$ & \\ 
\hline
\multirow{2}{*}{Stars+DM (SHMR)} & $8.38^{+3.62}_{-2.52}$ & $10.76^{+0.17}_{-0.17}$ & $<-33.63$ & $<0.32$ & $-$ & $-$ & $-$ & $21.85^{+1.18}_{-1.19}$ & $1.53^{+0.52}_{-0.53}$ & $1802.18^{+4.35}_{-4.31}$ & $-0.41^{+0.02}_{-0.01}$ \\ 
& $8.09^{+3.70}_{-2.36}$ & $10.76^{+0.16}_{-0.18}$ & $<-19.22$ & $-$ & $<0.07$ & $<0.39$ & $U$ & $21.85^{+1.17}_{-1.25}$ & $1.53^{+0.52}_{-0.52}$ & $1802.07^{+4.33}_{-4.33}$ & $0.23^{+0.02}_{-0.02}$ \\
\hline
\multirow{2}{*}{Stars+DM (NoSHMR)} & $23.71^{+8.75}_{-7.88}$ & $4.84^{+2.89}_{-2.93}$ & $<0.27$ & $-3.67^{+2.38}_{-3.26}$ & $-$ & $-$ & $-$ & $22.10^{+1.19}_{-1.16}$ & $1.79^{+0.48}_{-0.46}$ & $1803.98^{+2.68}_{-2.72}$ & $-0.15^{+0.01}_{-0.02}$ \\ 
 & $24.05^{+8.33}_{-7.58}$ & $4.48^{+2.60}_{-2.68}$ & $<-4.11$ & $-$ & $-3.61^{+2.03}_{-2.90}$ & $<0.12$ & $U$ & $22.10^{+1.18}_{-1.19}$ & $1.63^{+0.50}_{-0.48}$ & $1804.21^{+3.02}_{-3.02}$ & $-0.50^{+0.02}_{-0.01}$ \\
\tightdoublehline
\end{tabular}}
\end{minipage}
\end{table*}}}

{\renewcommand{\tabcolsep}{2.5mm}
{\renewcommand{\arraystretch}{2.5}
\begin{table*}
\begin{minipage}{\textwidth}
\centering
\caption{Results from the statistical analysis of NGC1052-DF4. considering NFW profile}\label{tab:results_DF4_NFW}
\resizebox*{\textwidth}{!}{
\begin{tabular}{c|ccc|ccccc|cc|c}
\tightdoublehline
& \multicolumn{10}{c}{{GR}}   \\
\hline
  & $c_{200}$ & $\log M_{200}$ & $\log L$ & $\beta_c$ & $\beta_0$ & $\beta_{\infty}$ & $r_a$ & $D$ & $\Upsilon_{\ast}$ & $v_{sys}$ & $\log\mathcal{B}^{i}_{j}$\\
  &  & M$_{\odot}$ & kpc &  &  &  & kpc & Mpc & & km s$^{-1}$ &  \\ 
\hline
\multirow{2}{*}{Stars + DM (SHMR)} 
 & $9.27^{+3.92}_{-2.85}$ & $10.74^{+0.16}_{-0.17}$ & $-$ & $<-0.11$ & $-$ & $-$ & $-$ & $21.86^{+1.24}_{-1.19}$ & $1.88^{+0.53}_{-0.53}$ & $1445.38^{+5.59}_{-5.36}$ & $-0.22^{+0.02}_{-0.02}$ \\
 & $8.84^{+4.08}_{-2.72}$ & $10.74^{+0.16}_{-0.17}$ & $-$ & $-$ & $<-0.60$ & $<-0.06$ & $U$ & $21.85^{+1.24}_{-1.20}$ & $1.90^{+0.53}_{-0.52}$ & $1445.28^{+6.22}_{-6.36}$ & $-0.76^{+0.02}_{-0.02}$\\
\hline
\multirow{2}{*}{Stars + DM (NoSHMR)} 
 & $24.09^{+8.44}_{-7.29}$ & $4.59^{+2.46}_{-2.76}$ & $-$ & $<-0.95$ & $-$ & $-$ & $-$ & $22.07^{+1.20}_{-1.16}$ & $1.92^{+0.50}_{-0.51}$ & $1445.98^{+2.35}_{-2.45}$ & $-0.13^{+0.02}_{-0.02}$ \\
 & $24.69^{+7.96}_{-7.46}$ & $4.12^{+2.54}_{-2.53}$ & $-$ & $-$ & $<-1.40$ & $<-0.24$ & $U$ & $22.01^{+1.21}_{-1.19}$ & $1.82^{+0.52}_{-0.51}$ & $1445.6^{+2.89}_{-2.92}$ & $-0.48^{+0.02}_{-0.01}$ \\
\hline
\multirow{1}{*}{Stars} 
 & $-$ & $-$ & $-$ & $<-0.99$ & $-$ & $-$ & $-$ & $22.04^{+1.23}_{-1.19}$ & $1.91^{+0.49}_{-0.48}$ & $1445.94^{+2.36}_{-2.38}$ & $0$ \\
\tightdoublehline
 & \multicolumn{10}{c}{{Non Minimal Coupling with $\epsilon = -1$}}   \\
\hline
  & $c_{200}$ & $\log M_{200}$ & $\log L$ & $\beta_c$ & $\beta_0$ & $\beta_{\infty}$ & $r_a$ & $D$ & $\Upsilon_{\ast}$ & $v_{sys}$ & $\log\mathcal{B}^{i}_{j}$ \\
  & & M$_{\odot}$  & kpc  &  &  &  & kpc & Mpc &  &  km s$^{-1}$ & \\ 
\hline
\multirow{2}{*}{Stars + DM (SHMR)} & $9.25^{+4.00}_{-2.79}$ & $10.75^{+0.15}_{-0.17}$ & $<-4.18$ & $<-0.17$ & $-$ & $-$ & $-$ & $21.92^{+1.18}_{-1.24}$ & $1.92^{+0.49}_{-0.52}$ & $1445.36^{+5.63}_{-5.55}$ & $-0.23^{+0.01}_{-0.02}$ \\ 
 & $9.05^{+3.95}_{-2.72}$ & $10.74^{+0.16}_{-0.17}$ & $<-0.23$ & $-$ & $<-0.03$ & $<0.38$ & $U$ & $21.91^{+1.20}_{-1.24}$ & $1.86^{+0.52}_{-0.52}$ & $1445.08^{+6.01}_{-6.15}$ & $-0.74^{+0.02}_{-0.02}$ \\
\hline
\multirow{2}{*}{Stars + DM (NoSHMR)} & $24.11^{+8.13}_{-7.32}$ & $4.35^{+2.61}_{-2.59}$ & $<-0.03$ & $<-0.16$ & $-$ & $-$ & $-$ & $21.98^{+1.23}_{-1.19}$ & $1.92^{+0.50}_{-0.50}$ & $1445.99^{+2.37}_{-2.41}$ & $-0.12^{+0.01}_{-0.01}$ \\ 
 & $24.46^{+8.22}_{-7.38}$ & $4.18^{+2.47}_{-2.55}$ & $<-4.19$ & $-$ & $<-0.59$ & $<0.28$ & $U$ & $22.00^{+1.19}_{-1.21}$ & $1.81^{+0.52}_{-0.50}$ & $1445.47^{+2.94}_{-2.93}$ & $-0.49^{+0.02}_{-0.01}$  \\
\tightdoublehline
 & \multicolumn{10}{c}{{Non Minimal Coupling with $\epsilon = +1$}}   \\
\hline
  & $c_{200}$ & $\log M_{200}$ & $\log L$ & $\beta_c$ & $\beta_0$ & $\beta_{\infty}$ & $r_a$ & $D$ & $\Upsilon_{\ast}$ & $v_{sys}$ & $\log\mathcal{B}^{i}_{j}$\\
  & & M$_{\odot}$  & kpc  &  &  &  & kpc & Mpc &  &  km s$^{-1}$ & \\ 
\hline
\multirow{2}{*}{Stars+DM (SHMR)} & $9.24^{+4.07}_{-2.81}$ & $10.76^{+0.16}_{-0.17}$ & $<-30.18$ & $<0.48$ & $-$ & $-$ & $-$ & $21.89^{+1.21}_{-1.16}$ & $1.89^{+0.52}_{-0.50}$ & $1445.41^{+5.51}_{-5.64}$ & $-0.22^{+0.01}_{-0.01}$  \\ 
& $8.64^{+3.89}_{-2.55}$ & $10.75^{+0.16}_{-0.17}$ & $<-4.91$ & $-$ & $<-0.02$ & $<0.40$ & $U$ & $21.91^{+1.21}_{-1.20}$ & $1.88^{+0.53}_{-0.51}$ & $1445.13^{+6.02}_{-6.29}$ & $-0.70^{+0.01}_{-0.01}$\\
\hline
\multirow{2}{*}{Stars+DM (NoSHMR)} & $24.07^{+8.03}_{-7.37}$ & $4.58^{+2.46}_{-2.76}$ & $<-38.48$ & $<-0.10$ & $-$ & $-$ & $-$ & $22.02^{+1.16}_{-1.15}$ & $1.90^{+0.51}_{-0.49}$ & $1445.99^{+2.43}_{-2.43}$ & $-0.12^{+0.02}_{-0.01}$ \\ 
 & $24.57^{+7.96}_{-7.64}$ & $4.27^{+2.40}_{-2.56}$ & $<1.80$ & $-$ & $<-0.60$ & $<0.29$ & $U$ & $22.01^{+1.20}_{-1.17}$ & $1.80^{+0.50}_{-0.50}$ & $1445.61^{+2.88}_{-3.04}$ & $-0.51^{+0.02}_{-0.01}$ \\
\tightdoublehline
\end{tabular}}
\end{minipage}
\end{table*}}}

{\renewcommand{\tabcolsep}{2.5mm}
{\renewcommand{\arraystretch}{2.5}
\begin{table*}
\begin{minipage}{\textwidth}
\centering
\caption{Results from the statistical analysis of Dragonfly 44 considering NFW profile}\label{tab:results_DF44_NFW}
\resizebox*{\textwidth}{!}{
\begin{tabular}{c|ccc|cccc|cc|c}
\tightdoublehline
& \multicolumn{10}{c}{GR}   \\
\hline
  & $c_{200}$ & $\log M_{200}$ & $\log L$ & $\beta_c$ & $\beta_0$ & $\beta_{\infty}$ & $r_a$ & $D$ & $\Upsilon_{\ast}$ & $\log\mathcal{B}^{i}_{j}$ \\
  &  & M$_{\odot}$ & kpc &  &  &  & kpc & Mpc &  &\\ 
\hline
\multirow{2}{*}{Stars + DM (SHMR)} 
 & $7.82^{+2.06}_{-1.63}$ & $10.90^{+0.15}_{-0.16}$ & $-$ & $-0.73^{+0.34}_{-0.50}$ & $-$ & $-$ & $-$ & $94.31^{+13.59}_{-13.51}$ & $1.49^{+0.37}_{-0.30}$ & $0$\\
& $7.54^{+3.15}_{-1.56}$ & $10.80^{+0.17}_{-0.18}$ & $-$ & $-$ & $<-1.21$ & $<-0.90$ & $U$ & $83.45^{+9.94}_{-10.93}$ & $1.45^{+0.36}_{-0.29}$ & $0.53^{+0.02}_{-0.01}$  \\
\hline
\multirow{2}{*}{Stars + DM (NoSHMR)} 
 & $12.37^{+6.14}_{-4.26}$ & $10.46^{+0.35}_{-0.24}$ & $-$ & $-0.92^{+0.42}_{-0.70}$ & $-$ & $-$ & $-$ & $100.81^{+14.22}_{-14.09}$ & $1.59^{+0.41}_{-0.33}$ & $0.19^{+0.01}_{-0.01}$ \\
& $17.72^{+12.32}_{-8.29}$ & $7.09^{+3.12}_{-5.35}$ & $-$ & $-$ & $U$ & $<-0.33$ & $U$ & $106.34^{+14.75}_{-15.28}$ & $1.94^{+0.48}_{-0.50}$ & $0.52^{+0.03}_{-0.02}$\\
\tightdoublehline
 & \multicolumn{10}{c}{Non Minimal Coupling with $\epsilon = -1$}   \\
\hline
  & $c_{200}$ & $\log M_{200}$ & $\log L$ & $\beta_c$ & $\beta_0$ & $\beta_{\infty}$ & $r_a$ & $D$ & $\Upsilon_{\ast}$ & $\log\mathcal{B}^{i}_{j}$ \\
  & & M$_{\odot}$  & kpc  &  &  &  & kpc & Mpc & & \\ 
\hline
\multirow{2}{*}{Stars + DM (SHMR)} & $7.70^{+1.99}_{-1.57}$ & $10.91^{+0.15}_{-0.15}$ & $<-2.08$ & $-0.72^{+0.34}_{-0.47}$ & $-$ & $-$ & $-$ & $94.84^{+13.47}_{-13.19}$ & $1.50^{+0.37}_{-0.31}$ & $0.01^{+0.02}_{-0.01}$ \\ 
 & $6.42^{+1.91}_{-1.50}$ & $10.87^{+0.15}_{-0.16}$ & $<-7.53$ & $-$ & $<-0.70$ & $<-0.44$ & $U$ & $89.15^{+13.78}_{-13.46}$ & $1.49^{+0.39}_{-0.31}$ & $0.50^{+0.01}_{-0.02}$ \\
\hline
\multirow{2}{*}{Stars + DM (NoSHMR)} & $13.07^{+6.61}_{-4.45}$ & $10.42^{+0.33}_{-0.24}$ & $<-12.5$ & $-0.96^{+0.44}_{-0.69}$ & $-$ & $-$ & $-$ & $101.04^{+13.89}_{-14.29}$ & $1.58^{+0.39}_{-0.32}$ & $0.20^{+0.01}_{-0.02}$\\ 
 & $12.85^{+12.99}_{-4.80}$ & $9.93^{+0.42}_{-7.48}$ & $<-3.35$ & $-$ & $<-1.38$ & $<-0.05$ & $U$ & $102.56^{+15.01}_{-15.06}$ & $1.75^{+0.50}_{-0.42}$ & $-0.25^{+0.02}_{-0.03}$\\
\tightdoublehline
 & \multicolumn{10}{c}{Non Minimal Coupling with $\epsilon = +1$}   \\
\hline
  & $c_{200}$ & $\log M_{200}$ & $\log L$ & $\beta_c$ & $\beta_0$ & $\beta_{\infty}$ & $r_a$ & $D$ & $\Upsilon_{\ast}$ & $\log\mathcal{B}^{i}_{j}$ \\
  & & M$_{\odot}$  & kpc  &  &  &  & kpc & Mpc & & \\ 
\hline
\multirow{2}{*}{Stars+DM (SHMR)} & $7.82^{+1.95}_{-1.65}$ & $10.90^{+0.14}_{-0.16}$ & $<-0.22$ & $-0.75^{+0.35}_{-0.47}$ & $-$ & $-$ & $-$ & $94.75^{+14.25}_{-13.62}$ & $1.52^{+0.39}_{-0.31}$ & $0.01^{+0.02}_{-0.02}$\\ 
 & $6.31^{+1.92}_{-1.45}$ & $10.87^{+0.15}_{-0.17}$ & $<-0.39$ & $-$ & $<-0.66$ & $<-0.39$ & $U$ & $91.10^{+14.43}_{-13.86}$ & $1.47^{+0.38}_{-0.29}$ & $0.52^{+0.02}_{-0.01}$ \\
\hline
\multirow{2}{*}{Stars+DM (NoSHMR)} & $12.25^{+6.33}_{-4.07}$ & $10.47^{+0.34}_{-0.27}$ & $<-0.41$ & $-0.91^{+0.40}_{-0.67}$ & $-$ & $-$ & $-$ & $100.56^{+14.26}_{-14.19}$ & $1.58^{+0.41}_{-0.32}$ & $0.24^{+0.02}_{-0.02}$\\ 
  & $11.28^{+5.16}_{-3.46}$ & $10.07^{+0.32}_{-0.37}$ & $<-0.37$ & $-$ & $<-1.28$ & $<-0.53$ & $U$ & $99.63^{+14.90}_{-14.03}$ & $1.60^{+0.43}_{-0.35}$ & $-0.12^{+0.02}_{-0.02}$ \\
\tightdoublehline
\end{tabular}}
\end{minipage}
\end{table*}}}

{\renewcommand{\tabcolsep}{2.5mm}
{\renewcommand{\arraystretch}{2.5}
\begin{table*}
\begin{minipage}{\textwidth}
\centering
\caption{Best-fit parameter sets for each halo profile, chosen from the runs with the highest $L$ values. For each profile, the three rows correspond to DF2, DF4, and DF44, respectively. The reported $L$ values are the maximum within the $1\sigma$ interval, and `U' denotes unconstrained parameters.}
\label{tab: All_Profiles}
\resizebox*{\textwidth}{!}{
\begin{tabular}{c|cccc|cccc|ccc|c|c}
\tightdoublehline
 & $c_{200}$ & $\log M_{200}$ & $\gamma$ & $\log L$ & $\beta_c$ & $\beta_0$ & $\beta_{\infty}$ & $r_a$ & $D$ & $\Upsilon_{\ast}$ & $v_{sys}$ & $\log\mathcal{B}^{i}_{j}$ & Case\\
 &  & M$_{\odot}$ &  & kpc  &  &  &  & kpc & Mpc &  &  km s$^{-1}$ &  &\\ 
\hline
\multirow{3}{*}{Burkert}
& $6.38^{+2.53}_{-1.68}$ & $10.82^{+0.16}_{-0.18}$ & $-$ & $<0.43$ & $-$ & $<-0.11$ & $<-0.29$ & $U$ & $21.75^{+1.21}_{-1.19}$ & $1.59^{+0.53}_{-0.51}$ & $1801.83^{+3.98}_{-3.88}$ & $0.28^{+0.03}_{-0.02}$ & SHMR+Radial ($\epsilon = -$) \\
& $24.75^{+8.14}_{-7.73}$ & $4.06^{+2.54}_{-2.50}$ & $-$ & $<-0.05$ & $-$ & $<-0.61$ & $<0.26$ & $U$ & $22.02^{+1.20}_{-1.21}$ & $1.81^{+0.51}_{-0.50}$ & $1445.57^{+2.85}_{-3.05}$ & $-0.68^{+0.02}_{-0.02}$ & NoSHMR+Radial ($\epsilon = -$) \\
& $>5.65$ & $>10.30$ & $-$ & $<0.10$ & $<0.18$ & $-$ & $-$ & $-$ & $102.16^{+14.47}_{-13.54}$ & $1.59^{+0.43}_{-0.35}$ & $-$ & $-0.80^{+0.02}_{-0.02}$ & NoSHMR+Const. ($\epsilon = -$) \\
\hline
\multirow{3}{*}{Hernquist}
& $20.63^{+9.67}_{-9.86}$ & $6.36^{+4.81}_{-3.79}$ & $-$ & $<3.61$ & $-3.82^{+2.44}_{-3.22}$ & $-$ & $-$ & $-$ & $22.15^{+1.20}_{-1.19}$ & $1.81^{+0.47}_{-0.46}$ & $1804.00^{+2.58}_{-2.72}$ & $-0.14^{+0.01}_{-0.02}$ & NoSHMR+Const. ($\epsilon = -$)\\
& $20.68^{+9.82}_{-9.83}$ & $6.33^{+4.92}_{-3.91}$ & $-$ & $<1.85$ & $<-0.12$ & $-$ & $-$ & $-$ & $22.03^{+1.18}_{-1.21}$ & $1.91^{+0.49}_{-0.49}$ & $1445.87^{+2.36}_{-2.34}$ & $-0.34^{+0.02}_{-0.01}$ & NoSHMR+Const. ($\epsilon = +$)\\
& $7.39^{+2.42}_{-1.84}$ & $11.16^{+0.15}_{-0.14}$ & $-$ & $<0.29$ & $>-9.43$ & $-$ & $-$ & $-$ & $108.73^{+14.38}_{-14.33}$ & $2.28^{+0.77}_{-0.55}$ & $-$ & $-0.14^{+0.01}_{-0.02}$ & SHMR+Const. ($\epsilon = +$)\\
\hline
\multirow{3}{*}{Dark-$\exp$-$\gamma$}
& $23.51^{+8.49}_{-7.67}$ & $4.78^{+2.91}_{-2.85}$ & $1.00^{+0.62}_{-0.60}$ & $<2.69$ & $-3.77^{+2.43}_{-3.13}$ & $-$ & $-$ & $-$ & $22.14^{+1.23}_{-1.22}$ & $1.79^{+0.48}_{-0.46}$ & $1804.02^{+2.66}_{-2.75}$ & $-0.23^{+0.01}_{-0.01}$&  NoSHMR+Const. ($\epsilon = -$)\\
& $7.82^{+3.70}_{-2.31}$ & $10.77^{+0.16}_{-0.17}$ & $0.74^{+0.63}_{-0.49}$ & $<0.49$ & $<0.42$ & $-$ & $-$ & $-$ & $21.87^{+1.21}_{-1.19}$ & $1.92^{+0.52}_{-0.52}$ & $1445.65^{+4.93}_{-5.06}$ & $-0.45^{+0.02}_{-0.02}$ & SHMR+Const. ($\epsilon = -$)\\
& $4.14^{+5.92}_{-2.36}$ & $12.95^{+5.03}_{-2.92}$ & $>0.11$ & $<0.59$ & $-$ & $-3.29^{+1.83}_{-3.09}$ & $-1.49^{+1.41}_{-2.78}$ & $U$ & $101.27^{+12.90}_{-13.99}$ & $1.58^{+0.40}_{-0.32}$ & $-$ & $-0.16^{+0.02}_{-0.02}$ & NoSHMR+Radial. ($\epsilon = -$)\\
\hline
\multirow{3}{*}{CNFW}
& $-$ & $-$ & $-$ & $-$ & $-$ & $-$ & $-$ & $-$ & $-$ & $-$ & $-$ & $-$ & $-$ \\
& $-$ & $-$ & $-$ & $-$ & $-$ & $-$ & $-$ & $-$ & $-$ & $-$ & $-$ & $-$ & $-$ \\
& $6.40^{+1.88}_{-1.56}$ & $10.86^{+0.16}_{-0.16}$ & $>24.77$ & $<-0.36$ & $-$ & $-1.67^{+1.02}_{-2.60}$ & $<-0.21$ & $U$ & $89.99^{+13.29}_{-13.88}$ & $1.48^{+0.37}_{-0.30}$ & $-$ & $0.54^{+0.02}_{-0.02}$ & SHMR+Radial ($\epsilon = -$) \\
\hline
\multirow{3}{*}{Einasto}
& $-$ & $-$ & $-$ & $-$ & $-$ & $-$ & $-$ & $-$ & $-$ & $-$ & $-$ & $-$ & $-$\\
& $-$ & $-$ & $-$ & $-$ & $-$ & $-$ & $-$ & $-$ & $-$ & $-$ & $-$ & $-$ & $-$\\
& $6.82^{+1.50}_{-0.92}$ & $10.94^{+0.17}_{-0.17}$ & $U$ & $<0.54$ & $<0.10$ & $-$ & $-$ & $-$ & $95.07^{+14.28}_{-13.35}$ & $1.54^{+0.39}_{-0.31}$ & $-$ & $0.50^{+0.02}_{-0.02}$ & SHMR+Const. ($\epsilon = -$)\\
\hline
\multirow{3}{*}{GPI}
& $-$ & $-$ & $-$ & $-$ & $-$ & $-$ & $-$ & $-$ & $-$ & $-$ & $-$ & $-$ & $-$\\
& $-$ & $-$ & $-$ & $-$ & $-$ & $-$ & $-$ & $-$ & $-$ & $-$ & $-$ & $-$ & $-$\\
& $3.79^{+1.33}_{-0.92}$ & $15.09^{+1.71}_{-2.34}$ & $22.15^{+16.79}_{-14.42}$ & $<2.76$ & $-$ & $-4.41^{+2.61}_{-2.70}$ & $<0.39$ & $U$ & $99.73^{+13.97}_{-14.49}$ & $1.60^{+0.45}_{-0.33}$ & $-$ & $1.79^{+0.05}_{-0.06}$ & NoSHMR+Radial ($\epsilon = -$)\\
\tightdoublehline
\end{tabular}}
\end{minipage}
\end{table*}}}

\bibliographystyle{apsrev4-1}
\bibliography{bibliography.bib}

\end{document}